\documentclass[onecolumn,floats,showpacs,prd,amssymb,floatfix,nofootinbib,balancelastpage,superscriptaddress,amsmath,showkeys]{revtex4}
\usepackage{epsfig}
\usepackage{latexsym}
\usepackage{graphicx}
\usepackage{amssymb}
\def\bea{\begin{eqnarray}}
\def\eea{\end{eqnarray}}
\def\be{\begin{equation}}
\def\ee{\end{equation}}

\def\b{\raisebox{14pt}{}\raisebox{-7pt}{}$\!\!$}

\newcommand{\msun}{M_{\odot}{\ }}
\newcommand{\ds}{{\sffamily DarkSUSY}}

\begin{document}

\title{Multi--wavelength signals of dark matter annihilations at the Galactic center}

\author{Marco Regis}
\email{regis@sissa.it} 

\author{Piero Ullio}
\email{ullio@sissa.it}

\affiliation{Scuola Internazionale Superiore di Studi Avanzati,
              Via Beirut 2-4, I-34014 Trieste, Italy and\\
              Istituto Nazionale di Fisica Nucleare,
              Sezione di Trieste, I-34014 Trieste, Italy}


\begin{abstract}
We perform a systematic study of the multi--wavelength signal induced by weakly interacting massive particle (WIMP) annihilations at the Galactic Center (GC).
Referring to a generic WIMP dark matter (DM) scenario and depending on astrophysical inputs, we discuss spectral and angular features and sketch correlations among signals in the different energy bands.
None of the components which have been associated to the GC source Sgr~A$^*$, nor the diffuse emission components from the GC region, have spectral or angular features typical of a DM source.
Still, data-sets at all energy bands, namely the radio, near infrared, X-ray and gamma-ray bands, contribute to place significant constraints on the WIMP parameter space. In general, the gamma-ray energy range is not the one with the largest signal to background ratio.
In the case of large magnetic fields close to the GC, X-ray data give the tightest bounds. The emission in the radio-band, which is less model dependent, is very constraining as well.
The recent detection by HESS of a GC gamma-ray source, and of a diffuse gamma-ray component, limits the possibility of a DM discovery with the next generation of gamma-ray telescopes, like GLAST and CTA.  We find that most of the region in the parameter space accessible to these instruments is actually already excluded at other wavelengths.
On the other hand, there may be still an open window to improve constraints with wide-field radio observations.

\end{abstract}




\maketitle

\section{Introduction}

Weakly interacting massive particles (WIMPs) are among the leading 
candidates for the dark matter (DM) component in the Universe. The framework is 
elegant and simple: stable WIMPs can be embedded in most extensions 
to the standard model of particle physics. In thermal equilibrium in the 
early Universe, they decouple from the primordial bath in the non-relativistic
regime. Their relic abundance scales approximatively with the inverse of their 
total pair annihilation rate into lighter particles: the weak-interaction coupling
ensures that, within the standard cosmological scenario, such relic density  
is of the order of the mean density of DM in the Universe today, as determined in 
cosmological observations (for comprehensive reviews on WIMP DM candidates
and their detection, see, e.g., \cite{Jungman:1995df,Bergstrom:2000pn,Bertone:2004pz})

In principle, one of the routes to test the hypothesis of WIMP DM stems from 
the bases of the framework themselves. Supposing that WIMPs are indeed the 
building blocks of all structures in the Universe, there is a (small but finite)
probability that WIMPs in DM halos, including the halo of 
the Milky Way, annihilate in pairs into detectable species. Indirect detection has mainly been focused on the search for 
a WIMP-induced component in the local antiproton, positron, and antideuteron 
cosmic-ray fluxes and for an excess in the high-energy gamma-ray galactic or extra-galactic flux (relevant constraints on the WIMP parameter space have been derived from such analyses; for recent results, see,  
e.g.,~\cite{Mack:2008wu,Wood:2008hx,Kachelriess:2007aj,Ahn:2007ty}). Since the 
gamma-ray signal scales with the square of the WIMP density along the line of 
sight, the Galactic center (GC) has been often indicated as the prime target.
In any self-consistent model for the distribution of DM in galactic halos, the DM
density is found to be maximal at the center of the system; numerical N-body 
simulations of hierarchical clustering in cold dark matter cosmologies find
configurations with very large overdensities, consistent with singular density 
profiles~\cite{Navarro:1996gj,Navarro:2003ew,Diemand:2005wv} 
(notice, however, that the simulations lack resolution to map the
distribution of DM on the very small scales which are relevant for WIMP signals;
at the same time, there is some tension between the simulation results and 
the DM profiles as derived from dynamical observations for certain classes of 
galaxies, see, e.g.~\cite{Flores:1994gz,Moore:1994yx,Gentile:2004tb}).

The Galactic center is an extraordinary site from several different points of 
view. Dynamical observations point to the presence of a supermassive black 
hole~\cite{Genzel:2000mj,Ghez:2000ay,Schodel:2002py},
with mass $M_{\rm BH} \sim 3 \times 10^6 \msun$, located very close to the 
dynamical center of the Galaxy, and most likely associated to the compact radio 
source labeled Sgr~A$^*$.  Infrared and X-ray counterparts have been identified 
for Sgr~A$^*$; GeV and TeV emissions in the direction of the GC have been 
detected as well, with the first data with high statistics and fair angular resolution
which have been obtained with the HESS air Cherenkov telescope~\cite{Aharonian:2004wa}.
Sgr~A$^*$ is an unusual source, certainly very different from
typical galactic or extragalactic compact sources associated to black holes. Most  
notably, under our perspective, it has a very low luminosity over the whole spectrum,
at a level at which it is plausible that a WIMP-induced component may be relevant. 

The emission associated to WIMP annihilations is expected to extend from the radio 
band up to gamma-ray frequencies. The peak of the gamma-ray luminosity stands at
the energy corresponding to a fraction (say one-third to one-twentieth) of the WIMP mass,
which is in turn in the few (tens of) GeV -- few TeV range; it is mostly associated to 
the chain of decays and/or hadronization processes initiated by two-body final state
particles from WIMP pair annihilations, leading to the production of neutral pions and their
subsequent decays into two photons. In an analogous chain, with analogous efficiency, 
high-energy electrons and positrons are produced by charged pions. Emitted in a region with large 
magnetic fields, as typical for accretion flows around supermassive black holes,
they give rise to synchrotron emission covering radio frequencies up to, possibly,
the X-ray band.

Numerous analyses have been dedicated to the study of the GC as a WIMP 
gamma-ray source, a list of recent references includes, 
e.g.,~\cite{Bergstrom:1997fj,Cesarini:2003nr,Hooper:2002ru,Hooper:2004vp,
Fornengo:2004kj,Horns:2004bk,Bergstrom:2004cy,
Profumo:2005xd,Zaharijas:2006qb,Aharonian:2006wh,
Dodelson:2007gd}. A prediction for the synchrotron emission has been discussed in 
Refs.~\cite{Gondolo:2000pn,Bertone:2001jv},  and refined on several aspects in 
Ref.~\cite{Aloisio:2004hy}; a comparison with X-ray data motivated by a class of heavy 
WIMP DM candidates is presented in Ref~\cite{Bergstrom:2006ny}.
We consider here the topic within  a self-consistent multi--wavelength approach.
Referring to a generic WIMP DM scenario, we discuss spectral and angular features, 
and  sketch the correlations among signals in the different energy bands. 
We illustrate which are the critical assumptions in deriving such conclusions, 
analyze them in the context of the currently available datasets, and make projections 
for the testability of the framework in the future. The procedure we follow is, to some
extent, analogous to the one adopted in Ref.~\cite{Colafrancesco:2005ji} for galaxy
clusters, and in Ref.~\cite{Colafrancesco:2006he} for dwarf satellites;  the list of relevant 
references include also~\cite{Colafrancesco:2000zv,Blasi:2002ct,Bertone:2002ms,
Tasitsiomi:2003vw,Baltz:2004bb,Hooper:2007kb}.

The paper is organized as follows: in Section~\ref{sec:data} we briefly summarize 
currently available observations on the GC. In Section~\ref{sec:source} we describe the DM source and the related mechanisms of photon production. In Section~\ref{sec:scalings} we compute the approximate scalings of the multi--wavelength spectrum, while in Section~\ref{sec:fulltreat} we perform the full treatment for some benchmark models. Then in Section~\ref{sec:result} we compare the DM--induced signal with the present limits and with the projected constraints of forthcoming experiments. Finally, Section~\ref{sec:conclusion} concludes.

\section{Overview of data on Sgr~A$^*$ and the GC region}
\label{sec:data}

The radio to sub-mm emission from Sgr~A$^*$ is characterized by a very hard spectrum:
the luminosity above $\nu \sim 1$~GHz scales approximately as $L_\nu \sim \nu^\alpha$ with 
$\alpha \simeq 0.8$ and cut-off at about $\nu \sim 10^3$~GHz (a compilation of 
available data and a full list of references is given, e.g.. in Ref.~\cite{Narayan:1997ku}).
We will show that such features do not seem to be compatible with the synchrotron 
emission induced by  WIMP annihilations, not even
with the observed flux reshaped by synchrotron self-absorption. In general, softer spectra 
are obtained, and the comparison with observations is useful to infer limits on the WIMP
parameter space. The tightest bound follows from the measurement at the lowest 
frequency, i.e. the upper bound on the flux density $S_\nu \le 0.05$~Jy at the frequency 
$\nu=408$~MHz, obtained with an interferometer with 4.3~arcsec angular acceptance 
at Jodrell Bank~\cite{davies76}.

\begin{figure}[t]
   \centering
   \includegraphics[width=10 cm]{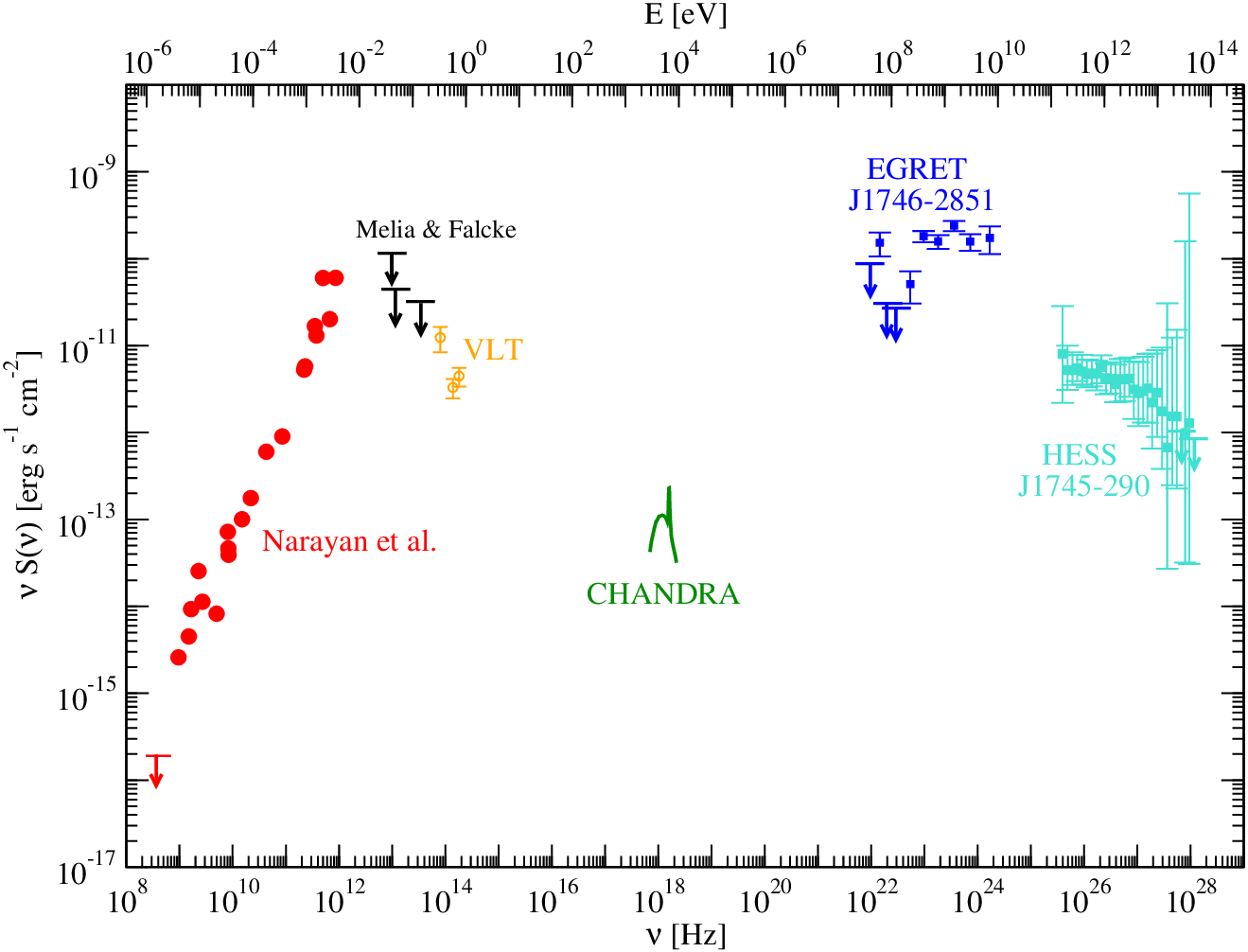}
\caption{Multi--wavelength spectrum of Sgr~A$^*$. The radio to X--ray emissions are shown in the quiescent state or at the epoch of lowest luminosity among available observations. The plotted $\gamma$--ray sources have positions compatible with Sgr~A$^*$; however, due to a poor angular resolution, EGRET cannot clearly identify the source and perhaps neither the HESS telescope.  See the text for details about the observations in each band.}
\label{fig:data}
 \end{figure}

Although variations are clearly seen in the radio flux density at different epochs, clean 
patterns of temporal dependencies have not been identified, see, e.g.,~\cite{Melia:2001dy};
the data plotted in Fig.~\ref{fig:data} are not time-averaged. At any given frequency,  
we show, among  the available measurements, the one corresponding to the epoch 
of lowest luminosity.

The angular size of the source depends on the frequency of observation.
At 1~GHz, a frequency at which it is expected that scattering in the interstellar medium
would wash out the true structure of Sgr~A$^*$,  it is of the order of 1.5 arcsec~\cite{davies76}.
At higher frequencies, the size shrinks proportionally to $\nu^{-2}$ up to  the measured 
value of 0.2~mas (about 1~AU in physical size) at about 86~GHz~\cite{Shen:2005cw}, 
possibly at the level of the intrinsic size of the source. We will show that, at radio frequencies, 
the angular size of  a WIMP-induced component is expected to be much larger than these
apparent or intrinsic angular dimensions; we find sizes for which it is actually interesting 
to compare with wide field images of the GC region rather than Sgr~A$^*$ alone. Among
the available surveys, we will refer again to those at the lowest frequency, namely at 90~cm.
An atlas of the diffuse radio emission in the Milky Way was presented in~\cite{Haslam:1982}. The evidence for a GC diffuse non-thermal source was enlighted in~\cite{LaRosa:2005ai}. Both of the maps have an angular resolution $\sim 1^{\circ}$, thus hiding the spatial structure of the diffuse emission in the innermost region. We will consider an image of the GC region constructed from VLA data, covering an area of $4 \times 5$ degrees and with angular resolution of 43~arcsec~\cite{LaRosa:2000}.

The near-infrared and X-ray emissions from Sgr~A$^*$ are characterized by a large variability  
(on different timescales in the two cases): quiescent values for the luminosity are plotted
in Fig.~\ref{fig:data}. 
The quiescent flux in the near-infrared has been recently detected 
with the VLT~\cite{Genzel:2003as,Clenet:2003tv}
as a point source with a position coincident with the supermassive black hole
within an accuracy of 10-20 mas, limited by faintness and by the proximity of one of the stars
orbiting the black hole~\cite{Schodel:2002py}. 
Launched in 1999, NASA's Chandra X--ray 
observatory is at present the most powerful X--ray detector, covering the energy range 
0.1~kev--10~kev with an angular resolution of 0.5 arcsec. During its observations, 
it has clearly discovered an X--ray source consistent with the position of 
Sgr~A$^*$ \cite{Baganoff:2001ju,Xu:2005et}, whose quiescent emission is well fitted by an absorbed thermal bremsstrahlung plus a Gaussian-line, plotted in Fig.~\ref{fig:data}. The spatial dimension of the X-ray source is $1.5$ arcsec. 
The process involving WIMP annihilations is expected to be steady, i.e. it cannot reproduce any time variability pattern. We will show that a X--ray flux at the quiescent level detected by Chandra can be obtained in the case of large WIMP densities and large magnetic field; moreover the source is predicted essentially as point--like, rather than the extended source seen by the Chandra detector.
We will use Sgr~A$^*$ infrared and X--ray data to set constraints on WIMP models.

Chandra detected also a diffuse emission in several regions within the inner 20 pc of the Galaxy. The reconstructed image covers a field of view of $17'\times 17'$ around Sgr~A$^*$ \cite{Muno:2004bs}. This diffuse emission could be consistently modeled as originating from a two--temperature diffuse plasma. The soft component ($kT\sim0.8$ keV) could be explained invoking different astrophysical mechanisms, while the origin of the hard component ($kT\gtrsim3$ keV), spatially uniform, is not clearly understood.
In principle it could be explained in terms of inverse Compton scattering on cosmic microwave background (CMB) induced by WIMP annihilations; however the detection of several emission lines and the inconsistency with limits at other frequencies make this hypothesis unplausible.

We come finally to gamma-ray observations. The EGRET team has reported the observation of 
a GC source in the energy range 100~MeV--20~GeV~\cite{MayerHasselwander:1998hg}. 
As it can be seen in Fig.~\ref{fig:data},
the luminosity of such a source exceeds by about one order of magnitude the luminosity
of Sgr~A$^*$ at any other frequency. The angular resolution of EGRET was rather poor, about
1~degree at 1~GeV, encompassing a large portion of the GC and not allowing for a clean 
identification of the emitter. In Ref.~\cite{Hooper:2002ru}, the authors argue that the improvement of the instrument angular resolution at multi-GeV energies should be taken into account in the 
data analysis, and conclude that it is actually possible to exclude the 
identification of the EGRET source with Sgr~A$^*$; in the same paper it is suggested that 
the comparison to set constraints on WIMP models should be with the diffuse background 
measured by EGRET in the GC region, rather than with the EGRET GC source.

The detection of TeV gamma-ray radiation from the GC has been reported by 
HESS~\cite{Aharonian:2004wa,Aharonian:2004jr,Aharonian:2006au}. Such a measurement 
has been confirmed, with a consistent spectrum, by MAGIC~\cite{Albert:2005kh} and 
supersedes previous results by CANGAROO~\cite{Tsuchiya:2004wv} and 
Whipple~\cite{Kosack:2004ri}, whose significantly different spectra is likely due to a
miscalibration of the detector and poorer statistics rather than variability of the source.
HESS has discovered a point source,  whose position is coincident with Sgr~A$^*$ within
7.3~arcsec~$\pm$~8.7~arcsec~(stat.)~$\pm$~8.5~arcsec~(syst.)~\cite{vanEldik:2007yi}, 
excluding the identification with the nearby supernova remnant  Sgr~A~East, but not with 
other candidates, such as a pulsar wind nebula recently discovered by 
Chandra~\cite{Wang:2005ya} which is only 8.7~arcsec away from Sgr~A$^*$.
The luminosity spectrum of the HESS point source is shown in Fig.~\ref{fig:data};
it is a rather features-less flux, $\phi_\gamma \sim E^{-\alpha}$ with spectral index
$\alpha \simeq 2.25$, extending from 160~GeV up to above 20~TeV. Even on the basis 
of the spectral characteristics only, without any consistency checks at other wavelength, 
it has been shown that it is rather implausible that such a source is due to WIMP 
annihilations only~\cite{Horns:2004bk,Bergstrom:2004cy,Profumo:2005xd,Aharonian:2006wh}.
HESS has also reported the detection of a diffuse gamma-ray emission along the 
central 300~pc of the GC ridge, within about 0.8~degree in longitude and 0.3~degree in 
latitude with respect to the GC. We will consider the central source and the diffuse emission
as maximal background level to understand the potential for a discovery of a WIMP component
with upcoming gamma-ray telescopes.

\section{DM WIMPs as a multi--wavelength source}
\label{sec:source}

The DM WIMP source scales with the number density of WIMP pairs locally in space, i.e. assuming a smooth (i.e. without substructures), spherically symmetric, and static dark matter distribution, with $\rho^2/2\,M_{\chi}^2$, with $\rho(r)$ being the halo mass density profile at the radius $r$, and $M_{\chi}$ the mass of the dark matter particle. Emitted stable species are nearly monochromatic if they are direct products of the annihilation (since the annihilating particles are essentially at rest); they have much broader spectra if they are generated in cascades with decays and/or hadronization processes of unstable two-body final states. For a given species $i$, the source function takes the form:
\be
Q_i(E,r)=(\sigma v)\,\frac{\rho(r)^2}{2\,M_{\chi}^2} \times \frac{dN_i}{dE}(E) \;,
\label{eqQ}
\ee
where $\sigma v$ is the annihilation rate at zero temperature, and $dN_i/dE$ is the number of particles $i$ emitted per annihilation in the energy interval $(E,E+dE)$, obtained by weighting spectra for single annihilation channels over the corresponding branching ratio.  The species which are relevant in a multi--wavelength analysis are photons, as well as electrons and positrons which act as sources for radiative processes. For most WIMP models, branching ratios for monochromatic emission in these channels are subdominant; in our analysis, we will concentrate on the components  with continuum spectra.
Prompt emission of photons proceeds mainly through the production and decay of neutral pions, hence the energy of these photons is in the gamma-ray band. For reference and to make transparent  the connection with the notation introduced below for radiative processes, the  $\gamma$--ray emissivity can be written as :
\be
j_{\gamma}(E,r)=Q_{\gamma}(E,r)\,E\;.
\ee 
In the energy range of interest for this analysis absorption is negligible, and fluxes or intensities can be straightforwardly derived summing contributions along the line of sight. E.g., the differential $\gamma$--ray flux is:
\be
\phi_{\gamma}(E,\theta)=\frac{1}{E}\int_{l.o.s.}ds\,\frac{j_{\gamma}(E,r(s,\theta))}{4\,\pi}
\label{eq:gammafl}
\ee
where the coordinate $s$ runs along the line of sight and $\theta$ is the angular off--set from the Galactic center.

\begin{figure}[t]
 \begin{minipage}[htb]{8cm}
   \centering
   \includegraphics[width=7.8cm]{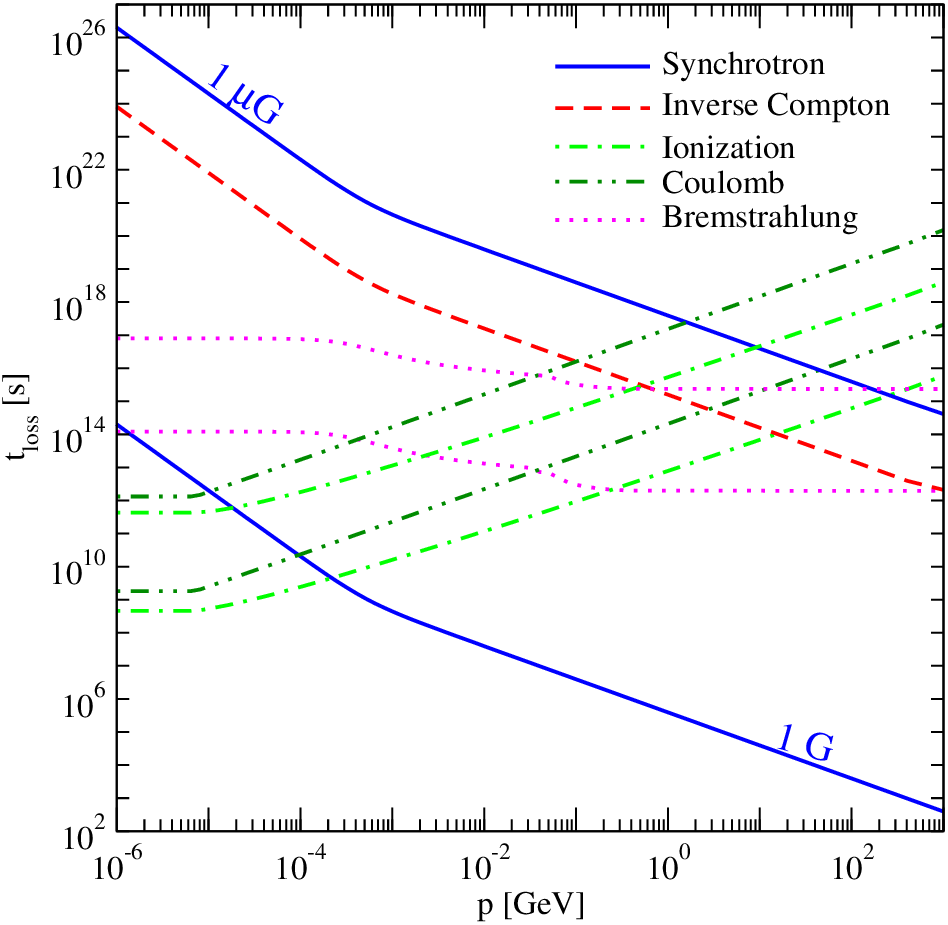}
 \end{minipage}
 \ \hspace{5mm} \
 \begin{minipage}[htb]{8cm}
   \centering
   \includegraphics[width=7.8cm]{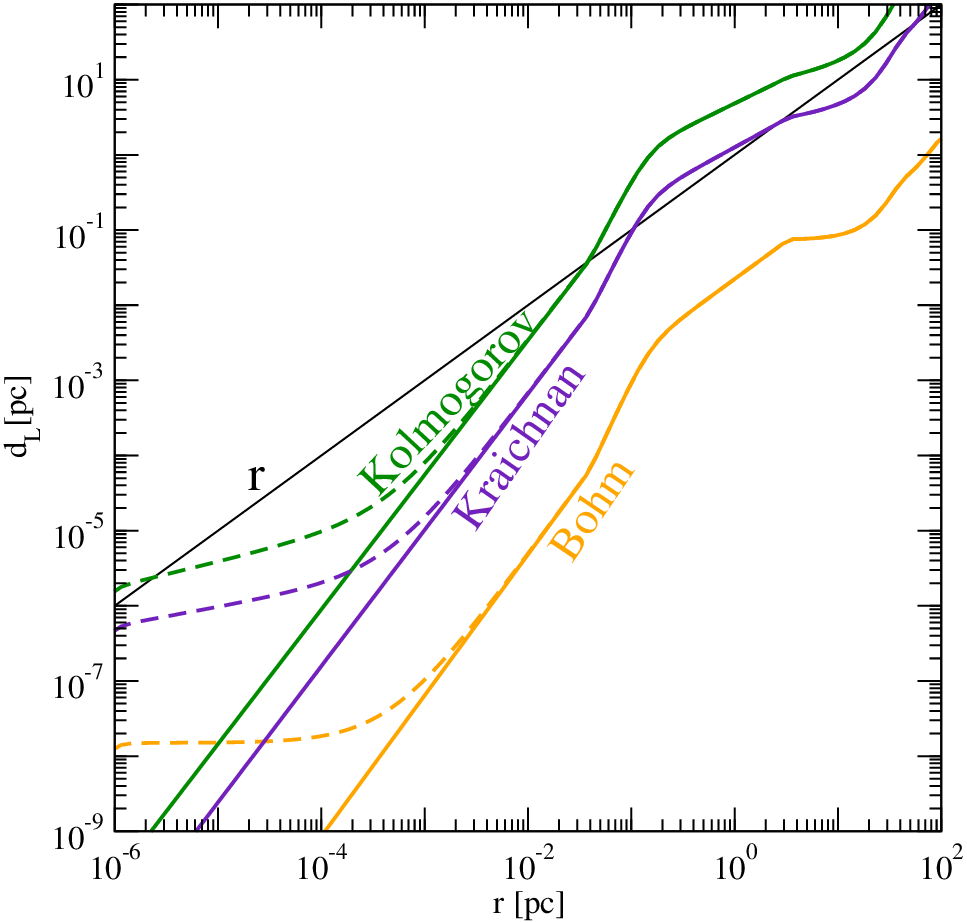}
 \end{minipage}
    \caption{{\it Left Panel:} Timescales for different radiative losses as a function of the $e^+ -e^-$ momentum. Synchrotron losses are shown for two reference values for the magnetic field: $B=1 \,\mu G, 1\,G$. Radiative losses associated to bremsstrahlung, ionization and Coulomb scattering are shown at the GC (lower curves) and at a distance of 100 pc from the GC (upper curves). {\it Right Panel:} Distance $d_L$ travelled by an electron with an injection energy of 1 GeV before losing most of its energy; three different guesses for the diffusion coefficient are shown, in the case of equipartition and reconnection magnetic field, see Fig.~\ref{fig:Bfield}a (same line styles).}
\label{fig:enloss}
 \end{figure} 

The emission through radiative losses involves charged particles, mainly electrons and positrons. Produced in WIMP pair annihilations, they propagate, losing and/or gaining energy. To describe this process, we consider the transport equation, in the limit of spherical symmetry, and for a stationary solution (see for example \cite{Strong:2007nh}; diffusive reacceleration is neglected):
\be
 -\frac{1}{r^2}\frac{\partial}{\partial r}\left[r^2 D\frac{\partial f}{\partial r} \right] 
  +v \frac{\partial f}{\partial r}-\frac{1}{3r^2}\frac{\partial}{\partial r}(r^2v)\,p\frac{\partial f}{\partial p}+\frac{1}{p^2}\frac{\partial}{\partial p}(\dot p p^2 f)=
  q( r, p)
\label{eqtransp}
\ee
where $f(r,p)$ is the $e^+-e^-$ distribution function at the equilibrium, at a given radius $r$ and in terms of the momentum $p$, related to the number density in the energy interval $(E,E+dE)$ by: $n_e(r,E)dE=4\pi \,p^2f(r,p)dp$; analogously,  for the WIMP source function of electrons or positrons, we have $Q_e(r,E)dE=4\pi \,p^2q(r,p)dp$. 
The first term on the left-hand side (l.h.s.) describes the spatial diffusion, with $D(r,p)$ being the diffusion coefficient. The second and third terms model an advective (convective) transport with an inflow (outflow) of the electrons and positrons toward (away from) the center of the system, being $v(r)$ the flow velocity of the medium. Finally, the last term on the l.h.s. describes the energy loss of due to radiative processes; $\dot p(r,p)=\sum_i dp_i(r,p)/dt$ is the sum of the rates of momentum loss associated to the radiative process $i$.

We apply Eq.~\ref{eqtransp} to the GC.
The radiative losses affecting the $e^+-e^-$ propagation are synchrotron emission, inverse Compton scattering on CMB and starlight, bremsstrahlung, ionization, and Coulomb scattering.
We model the galactic medium as composed by molecular (H$_2$), atomic (H{\footnotesize I}), and ionized (H{\footnotesize II}) gases. The density profiles are extracted from the description of the central molecular zone in~\cite{Ferriere:2007yq}, approximating their results under the assumptions of spherical symmetry.
The synchrotron loss rate is spatially dependent, scaling with the square of the local value of the magnetic field.
We plot in Fig.~\ref{fig:enloss}a the time--scale for the energy loss associated to each radiative process, defined as $t_{loss}=E/\dot E$. We show the synchrotron emission for two reference values of the magnetic field, while the two curves for bremsstrahlung, ionization and Coulomb scattering refer to the losses at the GC and at a distance of 100 pc from the GC.
We plot one curve for inverse Compton scattering; the time--scale is inversely proportional to the energy density of the background radiation; at the GC the energy density of the starlight component is considerably larger ($8$ eV cm$^{-3}$~\cite{Porter:2008ve}) with respect to the CMB ($0.25$ eV cm$^{-3}$). We sum the two components assuming a starlight energy density constant over the whole GC region.

\begin{figure}[t]
 \begin{minipage}[c]{8cm}
   \centering
   \includegraphics[width=7.8cm]{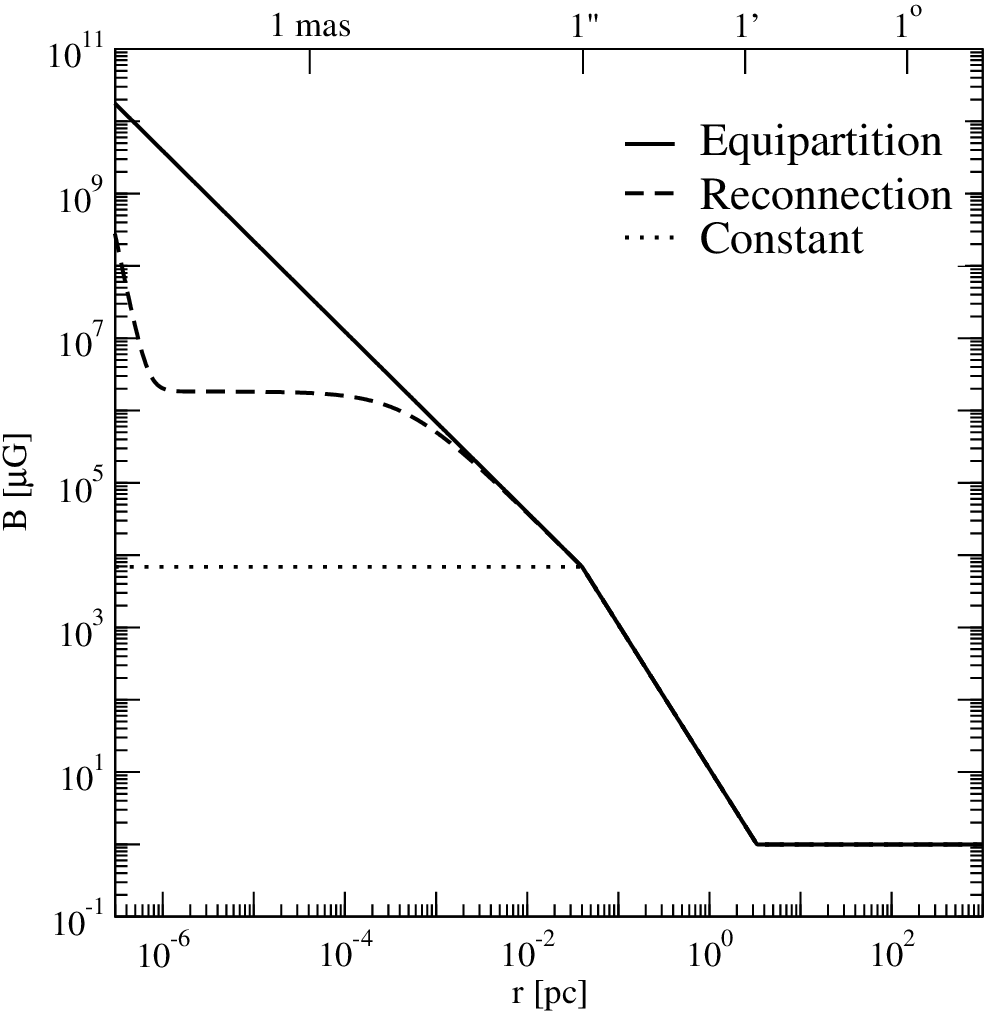}
 \end{minipage}
  \ \hspace{1mm} \
\begin{minipage}[c]{8cm}
   \centering
   \includegraphics[width=7.8cm]{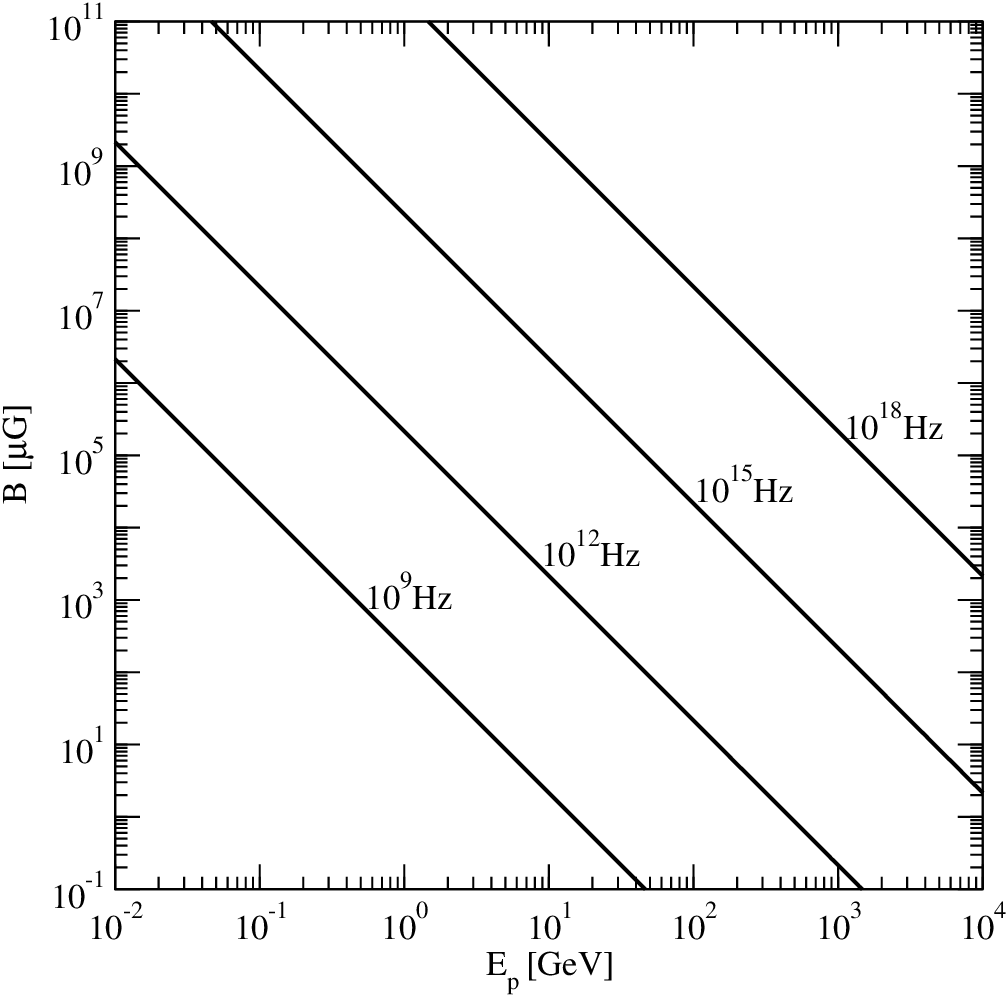}
 \end{minipage}
     \caption{{\it Left Panel:} Models for the magnetic fields in the central region of the Galaxy as a function of the distance from the GC. {\it Right Panel:} Magnetic field as a function of the synchrotron peak energy for few values of the observed frequencies.}
\label{fig:Bfield}
\end{figure}

The radial profile of the magnetic field is indeed an important ingredient in our analysis. 
Based on observations of nonthermal radio filaments, polarization of thermal dust emission, and synchrotron radiation from cosmic rays, the canonical picture of the Galactic center magnetosphere (for a review, see~\cite{Morris:2007jk}) describes the magnetic field with a dipolar geometry on large scale and as a pervasive field with strength of a mG throughout the central molecular zone (few hundred of pc). The recent discovery of a diffuse source of nonthermal synchrotron emission~\cite{LaRosa:2005ai} suggests, on the other hand, a mean magnetic field of order 10 $\mu$G on scales $\gtrsim$ few pc, unless reacceleration processes are invoked. It's important to note that such analyses constrain the mean magnetic field on scales $\gtrsim$ pc and do not exclude strong magnetic field in the innermost region.
Following again \cite{Melia:1992,Aloisio:2004hy}, we consider a magnetic field for the GC region satisfying the equipartition condition, namely, with the magnetic energy completely balancing the kinetic pressure:
\be
B(r)=3.9\,\cdot 10^4 \Big(\frac{0.01\,{\rm pc}}{r}\Big)^{5/4} \mu{\rm G}\;.
\label{eqBfield}
\ee
From a conservative point of view, this could be regarded as the maximal allowed magnetic field; we discuss this case together with two further possibilities: We follow \cite{Bergstrom:2006ny} and consider the case for a reduced magnetic field due to magnetic field line reconnection in turbulent plasma~\cite{Coker:1999gd}; as a toy model of an extreme case at the other hand with respect to the equipartition assumption, we allow also for a magnetic field which is constant within the accretion region. Outside the accretion region, assuming spherical infall and flux conservation, the magnetic field scales as $B\propto r^{-2}$ up to the large--scale value $B\simeq 1\,\mu$G \cite{Strong:2007nh}. The three different choices for the magnetic field radial profile are shown in Fig.~\ref{fig:Bfield}a.

Note that for magnetic fields B $\gtrsim 1$ G (as is typical for the innermost region of the Galaxy), the synchrotron losses dominate at all energies. For lower magnetic fields, i.e. at larger scales, inverse Compton scattering (and bremsstrahlung) becomes relevant in the ultra--relativistic regime, while ionization starts to dominate in the non--relativistic limit.

In order to estimate the relevance of spatial diffusion, we compare in Fig.~\ref{fig:enloss}b the physical scale $r$ with the distance diffused by electrons before losing most of their energy, $d_L\simeq (D\,E/\dot E)^{1/2}$.
In the quasilinear approximation of turbulent diffusion, the form of the diffusion coefficient $D$ can be expressed as $D(r,p)=1/3 r_g v_p (\delta B_{res} /B)^{-2}$, where $r_g=E/(e B)$ is the gyroradius of the electron, $v_p$ is the electron velocity, and $\delta B_{res}$ is the random component of the magnetic field at the resonant wavelength $k_{res}=1/r_g$.
On large scale (i.e., larger than about 100 pc) cosmic-ray data seem to indicate that the diffusion coefficient takes the form: $D=D_0\,(E_{GeV}/B_{\mu G})^{\alpha}$ with $\alpha \simeq 0.3-0.6$ and $D_0 \simeq 10^{27} - 10^{30} \rm{cm}^2\rm{s}^{-1}$~\cite{Strong:2007nh}; in the innermost region, the picture is much more uncertain.
Indirect constraints are derived in the models of~\cite{Aharonian:2004jr} and \cite{Ballantyne:2007tz}, when addressing the origin of the $\gamma$-ray source detected by HESS at the GC; in both analyses a significant reduction of the diffusion coefficient in the inner $10$ pc region is found.
On the modelling side, the relevance of diffusion is strictly connected with unknown variables needed in the description of turbulence, namely, the amplitude of the random magnetic field and the scale and the turbulence spectrum. As an example, one can assume comparable strengths for the regular and the random components of the magnetic field, and a power law, $ k^{-2+\alpha}$ for the turbulence spectrum.
For Bohm diffusion (typical when the coherence length of the magnetic field is comparable or greater with respect to the gyroradius of electrons), $\alpha =1$ and the coefficient reduces to $D=1/3 r_g v_p$; as one can see in Fig.~\ref{fig:enloss}b, in this case the effect of diffusion can be safely neglected.
Assuming a turbulent regime (in a homogeneous medium) with a scale of turbulence $\sim r$, we find that, for $\alpha=1/3$ (``Kolmogorov'', i.e. assuming a random flow of an incompressible fluid) and $\alpha=1/2$ (``Kraichnan'', which is more likely than the Kolmogorov spectrum in the case of the strong large-scale magnetic field), the diffusion can be relevant from the sub-pc scale in the first case, and it is marginally relevant around the pc scale in the second case.
Note the main ingredient here is the very large magnetic field considered in the innermost region of the Galaxy.
In our sample models, we find then that diffusion is either negligible over the whole central region or that it might be relevant only in its outer part, where, however, the DM source is expected to be less strong and have a less steep gradient than close to the central BH (see the discussion below). Therefore, we can foresee negligible to very mild effects from diffusion and, in what follows, for sake of simplicity and to make the discussion clearer, we will disregard the diffusion term.

We describe the accretion flow of gas onto the black hole in the innermost part of the Galaxy following \cite{Melia:1992,Aloisio:2004hy}, namely, we assume that the supersonic wind entering in the BH gravitational potential forms a bow shock dissipating kinetic energy and subsequently falls radially onto the BH. We take a spherical accretion and a steady flow, and estimate the region of the accretion as $R_{acc}=2GM/v_{flow}^2$, where $v_{flow}\simeq 500-700$ km s$^{-1}$ is the Galactic wind velocity and thus $R_{acc}\sim 0.04$~pc~\cite{Melia:1992}.
The radial infall velocity of the gas is
\be
v(r)=-c\sqrt{\frac{R_{BH}}{r}}
\label{eqgasvel}
\ee
A particle propagating in such accretion flow gains momentum since it feels an adiabatic compression in the BH direction.

The Galactic center lobe is a radio continuum emission spanning the central degree of the Galaxy with a bipolar structure.
Recent mid-infrared observations~\cite{BlandHawthorn:2002ij} suggest the idea that the emission associated to the GC lobe is a sign of a GC outflow, in particular, a starburst outflow. The associated large-scale bipolar wind could affect the transport equation Eq.~\ref{eqtransp}, convecting electrons and positrons. Assuming a velocity $\sim 10^2$ km/s~\cite{BlandHawthorn:2002ij}, this effect is negligible in the innermost region, while it can be relevant on larger scales. On the other hand, although the model of~\cite{BlandHawthorn:2002ij} is probably the most intriguing, one can resort to other mechanisms explaining the origin of the Galactic center lobe (for a recent review, see, e.g.,~\cite{Law:2007zh}). In the following we choose to neglect the effect of such a possible wind.

The solution of Eq.~\ref{eqtransp} provides the $e^+/e^-$ number density $n_e$ in the stationary limit.  For a radiative process $i$, with associated power $P_{i}$, the photon emissivity is given by folding $n_e$ with the power \cite{Rybicki}:
\be
j_{i}(\nu,r)=2\int^{M_{\chi}}_{m_e}dE\, P_{i}(r,E,\nu)\, n_e(r,E)\;,
\label{eqjsynch}
\ee
where the factor $2$ takes into account electrons and positrons (in WIMP annihilations, as well as during propagation, there is perfect symmetry between particles and antiparticles).
For any given emission mechanism, the associated luminosity at frequency $\nu$ is
\be
L_i(\nu)=\int d^3r\,j_{i}(\nu,r)\;,
\label{eqlumin}
\ee
while the intensity measured by a detector can be estimated as
\be
S_{i}(\nu,\theta,\theta_d) =\int d\Omega' \exp\Big(-\frac{\tan^2\theta'}{2\,\tan^2\theta_d}\Big)\int_{l.o.s.}dI_{i}(\nu,s,\tilde \theta)\;.
\label{eqIsynch}
\ee
Here $\theta$ labels the direction of observation, i.e. the angular off--set with respect to the GC and we are performing an angular integral assuming a circular Gaussian resolution of width  $\theta_d$ for the detector. 
$dI_i$ is the differential of the intensity of radiation $I_i$: within the increment $ds$ along a line of sight, there is a gain in intensity $j_i/(4\,\pi)\,ds$, while a decrease $\alpha\,I_i\,ds$ could be due to absorption.
$I_i$ follows from the solution of the differential equation:
\be
\frac{dI_{i}(\nu,s,\tilde \theta)}{ds}=-\alpha(\nu,s,\tilde \theta)\,I_i(\nu,s,\tilde \theta)+\frac{j_i(\nu,s,\tilde \theta)}{4\pi}
\label{eqdiffIsynch}
\ee
where $\tilde \theta$ is the angular off-set from the GC of the line of sight along which $I_i$ is calculated, as selected by $\theta$ and the angular variables of integration $\theta'$ and $\phi'$.
If absorption is negligible, the second integrand of Eq.~\ref{eqIsynch} reduces to $dI_{i}(\nu,s,\tilde \theta)=ds\,j_{i}(\nu,s,\tilde \theta)/(4\,\pi)$.

At  low and intermediate frequencies, i.e. in the radio band up to (possibly) the X-ray band, the DM signal is mostly due to  synchrotron radiation. The power for synchrotron emission takes the form \cite{Rybicki}:
\be
P_{syn} (r,E,\nu)= \frac{\sqrt{3}\,e^3}{m_e c^2} \,B(r) F(\nu/\nu_c)\;,
\ee
where $m_e$ is the electron mass, the critical synchrotron frequency is defined as $\nu_c \equiv  3/(4\,\pi) \cdot {c\,e}/{(m_e c^2)^3} B(r) E^2$,  and $F(t) \equiv t \int_t^\infty dz K_{5/3}(z)$ is the standard function setting the spectral behavior of synchrotron radiation. At radio frequencies, there are configurations for the parameters in the model for which synchrotron self-absorption is a relevant effect~\cite{Gondolo:2000pn,Bertone:2001jv,Aloisio:2004hy}; we include it in our analysis implementing the standard form for $\alpha$ in this case. In~\cite{Aloisio:2004hy} it was shown that, on the other hand, we can safely neglect synchrotron self-Compton effects. For the signal in the UV and soft--X band, we need to take into account  the photoelectric effect on the interstellar dust; this is an effect taking place mostly outside the region of emission hence we can model it a posteriori through an attenuation factor.

The emission through inverse Compton (IC) scattering of the ultra--relativistic electrons from WIMP annihilations on cosmic microwave or starlight background photons, could be relevant as well,
especially when targeting the GC region rather than the Sgr~A$^*$ only. This emission spans the
X-band up to the (soft) $\gamma$-ray band. The inverse Compton power is given by
\be
P_{IC}(r,E,\nu) = c\,h\nu \int d\epsilon\, \frac{dn_\gamma}{d\epsilon}(\epsilon,r)\,\sigma(\epsilon,\nu,E) 
 \label{eqPIC}
\ee
where $\epsilon$ is the energy of the target photons, $dn_\gamma/d\epsilon$ is their differential energy
spectrum, and  $\sigma$ is the Klein--Nishina cross section. Finally, a very faint emission is expected
in case of bremsstrahlung, ionization, and Coulomb scattering; we will not consider them in our analysis.

\section{The multi--wavelength seed in an approximate approach}
\label{sec:scalings}

\begin{figure}[t]
 \begin{minipage}[htb]{8cm}
   \centering
   \includegraphics[width=7.8cm]{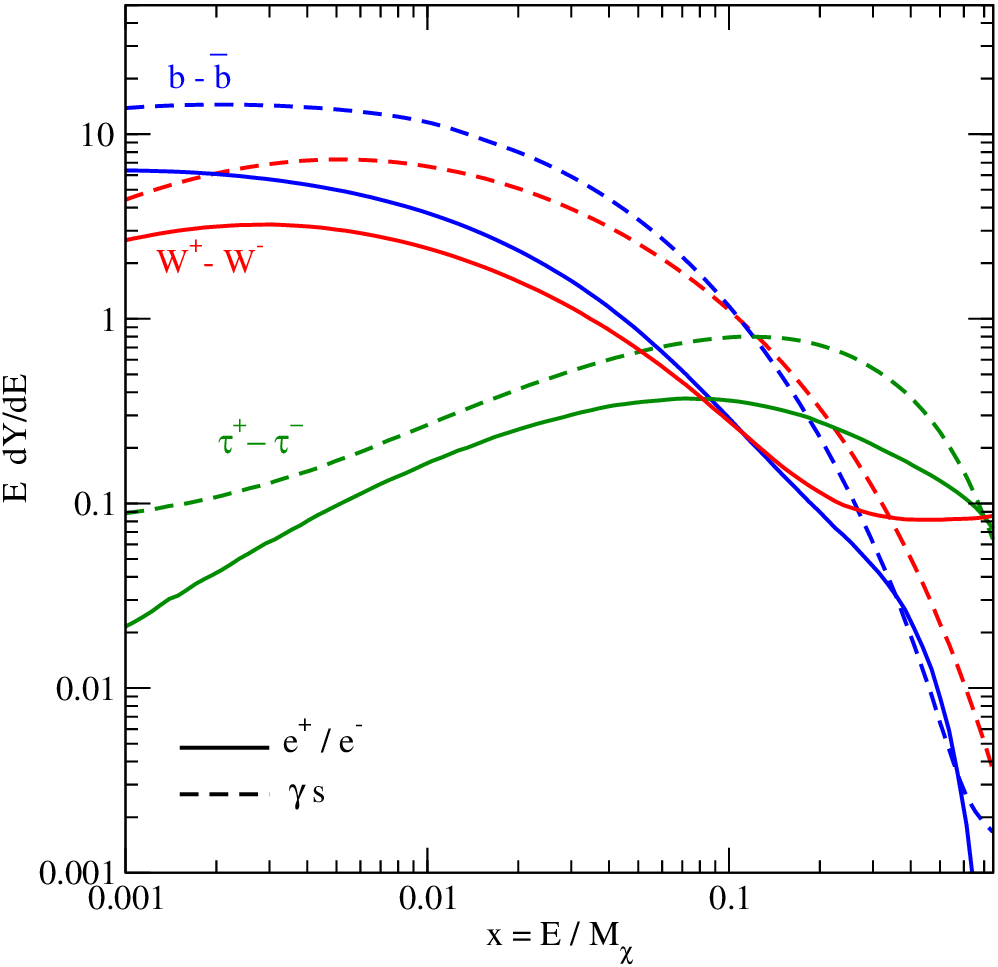}
 \end{minipage}
 \ \hspace{1mm} \
 \begin{minipage}[htb]{8cm}
   \centering
   \includegraphics[width=7.8cm]{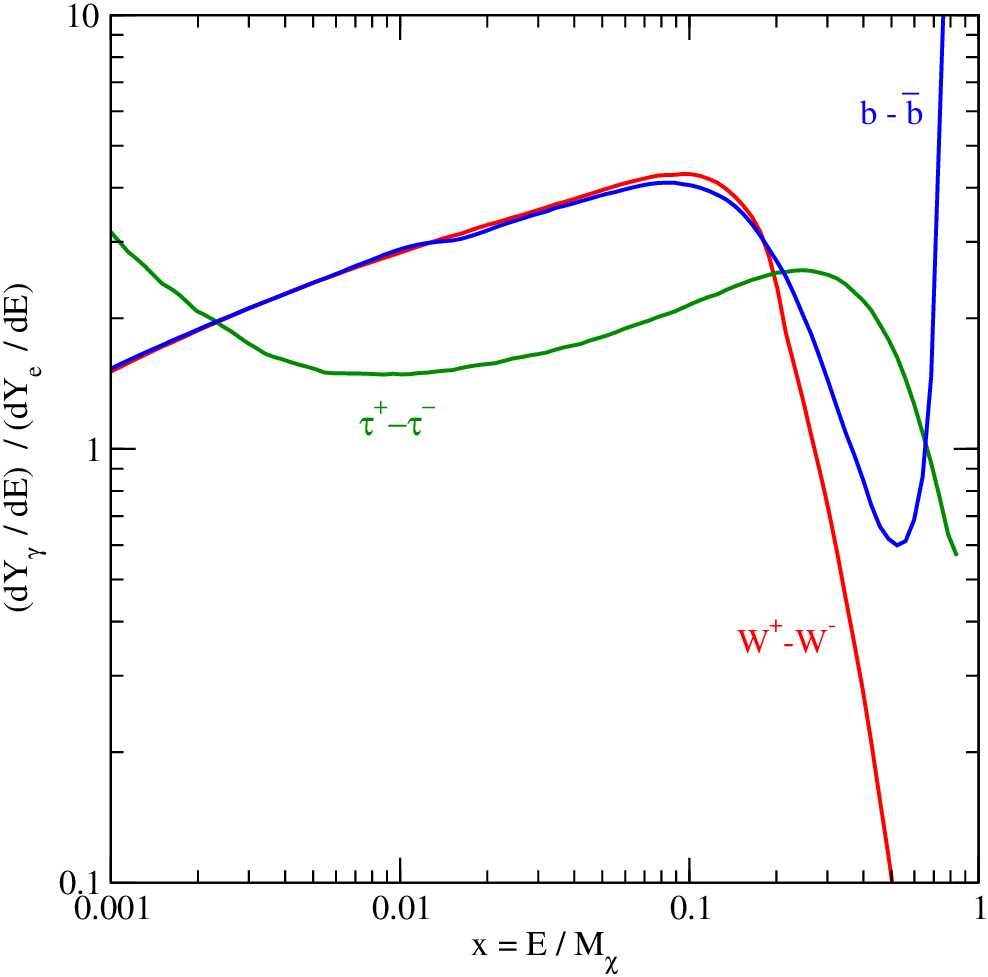}
 \end{minipage}
    \caption{{\it Left Panel:} $\gamma$--ray and $e^+-e^-$ spectra per annihilation for a $1$ TeV WIMP. The three annihilation channels $b -\bar b$, $W^+-W^-$, and $\tau^+-\tau^-$ are taken as references. {\it Right Panel:} Multiplicity between the electron and photon yields $dN_{\gamma}/dE \times (dN_e/dE)^{-1}$ for a $1$ TeV WIMP with the same annihilation modes as in the left panel.}
\label{fig:yield}
 \end{figure} 

In this section we sketch in a simple and analytic form the scalings of the dark matter induced signal depending on various assumptions in the model.
Eq.~\ref{eqtransp} does not admit in general an analytic solution. However, when the radiative loss term dominates (and thus the first three terms are negligible), one finds simply:
\be
n_e(r,E)=\frac{1}{\dot E (r,E)}\int^{M_{\chi}}_{E}dE\, Q_e(E,r)
\label{eqdnde}
\ee
where $\dot E$ comes from $\dot p$ in Eq.~\ref{eqtransp} mapping momentum into energy.
We have already stressed that synchrotron processes are the main effect for energy losses and radiative emissivity. 
We can focus, for the moment, on this mechanism, and write the energy loss rate as $\dot E = \dot E_{syn}= 4/9 \cdot (c\,e^4)/(m_ec^2)^4 B(r)^2 E^2$, and the induced synchrotron luminosity as
\be
\nu L_{\nu}^{syn}=\,4\pi \nu \frac{\sigma v}{M_{\chi}^2}\int dr\, r^2 \rho(r)^2 \int^{M_{\chi}}_E \frac{P_{syn}(\nu,r,E)}{\dot E_{syn}(r,E)} Y_e(E) =\,\frac{9\sqrt{3}}{4} \frac{\sigma v}{M_{\chi}^2}\int dr \,r^2 \rho(r)^2  E_p \, Y_e(E_p)\\
\label{eq:nuLnus}
\ee
where we have defined $Y_e(E)=\int^{M_{\chi}}_E dE' dN_e/dE'$, and in the last step we have implemented the monochromatic approximation for the synchrotron power, i.e. assumed  $F(\nu/\nu_c)\sim \delta(\nu/\nu_c-0.29)$ \cite{Rybicki}. In the monochromatic approximation there is a one-to-one correspondence between the energy of the radiating electron (peak energy in the power) and the frequency of the emitted photon, that is $E_p=\nu^{{1}/{2}}(0.29\, B(r)\,c_0)^{-{1}/{2}}$ with $c_0=3/(4\,\pi) \cdot {c\,e}/{(m_e c^2)^3}$, or, introducing values for numerical constants, the peak energy in GeV is $\widehat{E}_{p}\simeq 0.463\,\widehat{\nu}^{{1}/{2}}\widehat{B}^{-{1}/{2}}$, with $\widehat{\nu}$ the frequency in ~GHz and $\widehat{B}$ the magnetic field in~mG. Analogously, the induced $\gamma$--ray luminosity is
\be
\nu L_{\nu}^{\gamma}=\,2\pi\, \frac{\sigma v}{M_{\chi}^2}\int dr\, r^2 \rho(r)^2\,E^2\,\frac{dN_{\gamma}}{dE} \;.
\label{eq:nuLnug}
\ee

It is useful to make a few simple guesses on some of the quantities introduced above. 
Along the line of \cite{Bergstrom:1997fj}, we assume the $\gamma$--ray spectrum per annihilation following the law: $dN_{\gamma}/dx \simeq \tilde A\, x^{-\tilde B}e^{-\tilde C x}$, with $x\equiv E/M_{\chi}$.
It is also a fair assumption to approximate the integrated $e^+-e^-$ yield as a power law plus an exponential cutoff: $Y_e(E)\simeq A\,x^{-B}e^{-Cx}$.
The differential yields of secondary photons and $e^+-e^-$ are plotted in Fig.~\ref{fig:yield}a, for three sample cases of two-body final states from WIMP pair annihilations. These plots are obtained linking to simulations of decay/hadronization performed with the PYTHIA Monte--Carlo package \cite{Sjostrand:1993yb} and stored libraries contained in the \ds~package~\cite{Gondolo:2004sc}; we will refer to such kind of simulations everywhere in the paper when making detailed estimates of WIMP induced signals.
As the simplest guess for radial dependence  for the magnetic field and the DM profile, we consider the single power-law scalings,  $B(r)=B_0(r/r_0)^{-\beta}$ and $\rho(r)=\rho_0(r/a)^{-\gamma}$. Eqs.~\ref{eq:nuLnus} and \ref{eq:nuLnug} become:
\be
\left\{
\begin{split}
\nu L_{\nu}^{syn}=&\frac{1.8\,A}{0.463^B} \frac{\sigma v}{M_{\chi}^2} \rho_0^2\,a^{2\gamma} 
\frac{\left(\widehat{\nu}/\widehat{B}_{0}\right)^{(1-B)/2}}{\widehat{M}_{\chi}^{-B}} 
 \, \int dr\, r^{2-2\gamma}\left(\frac{r}{r_0}\right)^{\frac{\beta}{2}(1-B)}
 exp\left[-\frac{C}{\sqrt{4.66}} \frac{\left(\widehat{\nu}/\widehat{B}_{0}\right)^{1/2}} 
 {\widehat{M}_{\chi}}\left(\frac{r}{r_0}\right)^{\frac{\beta}{2}}\right]
\,{\rm GeV} \\
\nu L_{\nu}^{\gamma}=&2\pi\tilde A \frac{\sigma v}{M_{\chi}^2} \rho_0^2\,a^{2\gamma} 
\frac{\widehat{E}^{2-\tilde B}}{\widehat{M}_{\chi}^{1-\tilde B}}
\int dr\, r^{2-2\gamma}exp\left[-\tilde C \frac{\widehat{E}}{\widehat{M}_{\chi}}\right]
\,{\rm GeV}
\end{split}
\right.
\label{eq:nuLnusg}
\ee
with $\widehat{M}_{\chi}$ the WIMP mass in GeV.

The right-hand-sides of Eq.~\ref{eq:nuLnusg} show some differences. 
For the gamma-ray luminosity, the energy cutoff follows simply from energy conservation and thus scales with the dark matter mass, except for a O(1) factor  related to the annihilation mode. For synchrotron emission, at a fixed mass, the frequency cutoff  increases with the magnetic field, again except for a O(1) factor related to the annihilation channel.
Away from the cutoff, the synchrotron emissivity tends to originate from a larger spatial region with respect to the $\gamma$--ray case, due to the additional positive power $\beta/2 (1-B)$ in the radial dependence.
At fixed mass and frequency, if the magnetic field is large enough to avoid the frequency cutoff, the synchrotron signal is wider than the gamma-ray signal.  This is typically the case in the
radio band and, to a much smaller extent, in the infrared band.
Going to very high observed frequencies, however, the magnetic field (or the energy of the radiating  electron or positron) needs to increase to exceedingly large values, which might be met only very
close to the central BH (or for extremely massive WIMPs and/or hard $e^+-e^-$ spectrum, as encoded in the factor $C$ of Eq.~\ref{eq:nuLnusg}). 
Scalings of the required magnetic field, as a function of peak radiating energy, for a few values of the observed frequency are shown in Fig.~\ref{fig:Bfield}b:
one can see that for the observed frequencies getting into the X-ray band (say $10^{18}$~Hz) 
a very small radial interval is selected, corresponding to the largest allowed value for the magnetic 
field. Hence, in this case the synchrotron signal is actually expected to be originated in a very small
region around the central BH, possibly much smaller compared to the gamma-ray flux.

We can now make a sketchy estimate to find which of the limits in the different bands in Fig.~\ref{fig:data} might be more constraining. We write the ratio between synchrotron and gamma-ray luminosity in the form:
\be
r=\frac{\nu L_{\nu}^{syn}}{\nu L_{\nu}^{\gamma}}
= \frac{1.8}{2\pi\,0.463^B} 
 \frac{A}{\tilde A} 
\frac{\widehat{M}_{\chi}^{1+B-\tilde B}\,\widehat{\nu}^{(1-B)/2}}{\widehat{E}^{2-\tilde B}}
\frac{\int dr\, r^{2-2\gamma} \left[\widehat{B}(r)\right]^{-(1-B)/2}
\exp\left[-\frac{C E_p(r) -\tilde C E}{{M}_{\chi}}\right]}
{\int dr\, r^{2-2\gamma}}\;.
\label{eq:nuLfrac}
\ee

\begin{figure}[t]
   \centering
   \includegraphics[width=10 cm]{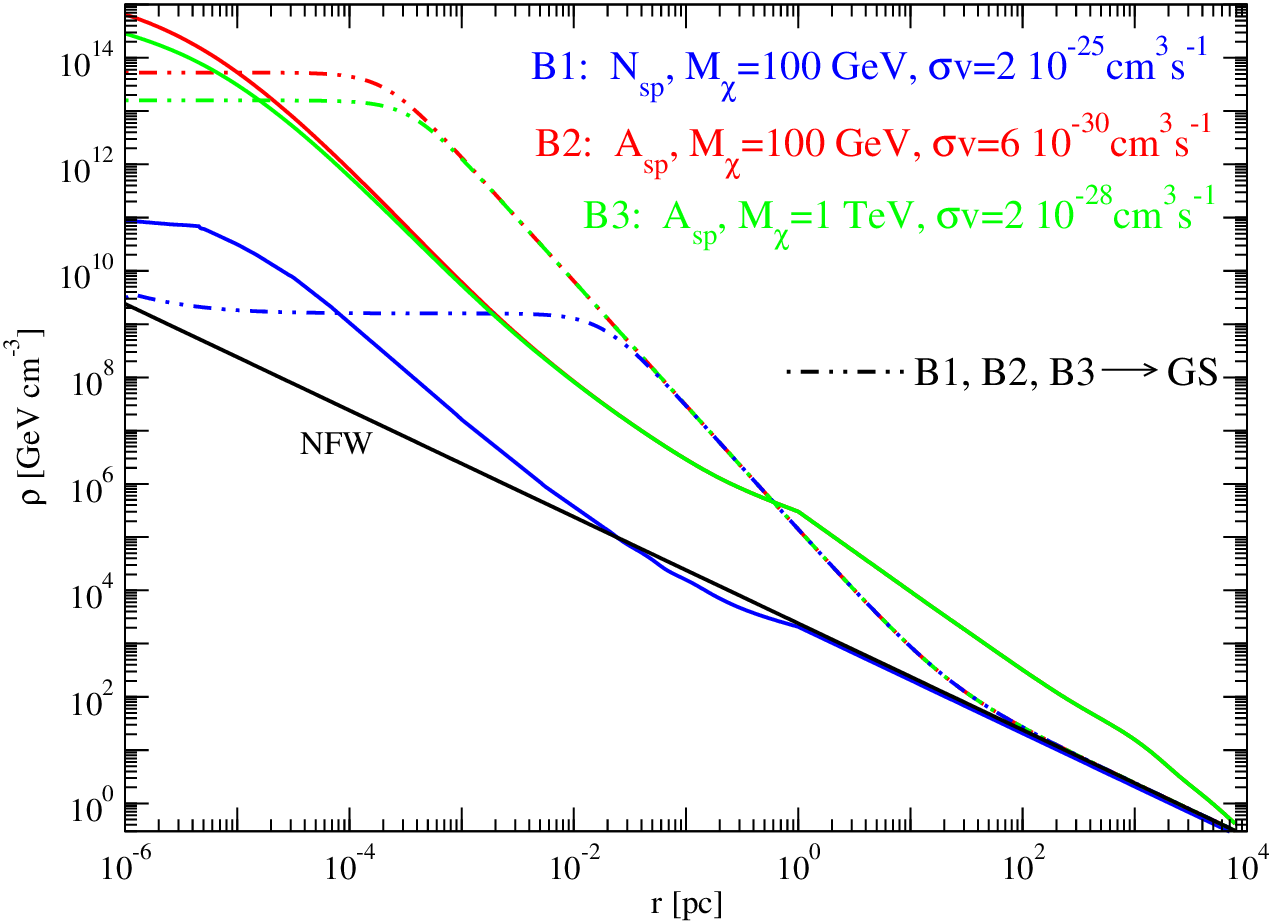}
    \caption{Dark matter profiles for the benchmark models B1, B2, and B3. For comparison we plot also the NFW profile and NFW profiles modified by the original prescription by Gondolo-Silk (GS) to account for the growth of the central black hole: the value of the ratio $(\sigma v)/M_{\chi}$ are the same as in the benchmark models.}
\label{figprofiles}
\end{figure}

In Fig.~\ref{fig:yield}b we plot the relative multiplicity between photons and electrons for the three benchmark final states from WIMP pair annihilations considered in  Fig.~\ref{fig:yield}a. This 
illustrates the fact that, sufficiently far away from the energy cutoff and for a generic WIMP annihilation 
channel (except, of course, for the case of prompt emission of monochromatic gammas, and/or electrons/positrons we are not considering here), the photon and electron/positron yields are comparable and hence that it is difficult to avoid the correlation between the gamma and the synchrotron signals by selecting a specific WIMP model. In  Eq.~\ref{eq:nuLfrac} this implies that the ratio $A/\tilde A$ is typically O(1). 
The last term in Eq.~\ref{eq:nuLfrac} does critically enter in boosting or suppressing the ratio
of luminosities only in case the exponential cutoff (or the upper limit in the radial integral) is playing 
a role, i.e. at very large observational frequencies for synchrotron emission (the X-ray band) or
for shallow density profiles.
Restricting to the case of singular halo profiles, and, e.g. the radio band, it is of order O(1) or O(0.1). To see this more precisely, let us take $W^+-W^-$ as the annihilation channel, as an intermediate case between the soft quark spectra and the hard leptonic spectra. We find that integrated $e^+-e^-$ yield, for masses in the range 
$M_{\chi}$ = 100 GeV--10 TeV, can be fairly well  approximated by  $(A,B,C) \simeq(0.1,0.7,3)$;
the differential $\gamma$--spectrum was fitted in~\cite{Bergstrom:1997fj}, finding 
$(\tilde A,\tilde B,\tilde C)\simeq(0.73,1.5,7.8)$. Since $B\simeq0.7$, the rescaling factor $\left[\widehat{B}(r)\right]^{-(1-B)/2}$, in the integral in the numerator of Eq.~\ref{eq:nuLfrac}, varies at most 
between about 0.09 and 3, hence we can assume as a sample value for the suppression expected from the ratio of integrals a factor of 0.5. Inserting all fit parameters in Eq.~\ref{eq:nuLfrac}, we get:
\be
r_W \sim \,9\cdot 10^{-2} 
\left(\frac{M_\chi}{100\,{\rm GeV}}\right)^{0.2} 
\left(\frac{\nu}{1\,{\rm GHz}}\right)^{0.15}
\left(\frac{1\,{\rm GeV}}{E}\right)^{0.5} 
\label{eq:fracW}
\ee
We find hence that the radio and $\gamma$--ray luminosities are at a comparable level, while as it can be seen in Fig.~\ref{fig:data},  constraints in the $\gamma$--ray band are several orders of magnitude weaker than at radio wavelengths. Although the luminosities of Eqs.~\ref{eq:nuLnus} and ~\ref{eq:nuLnug} cannot be directly compared with such experimental data, since they are integrated over the whole emission region, which can be significantly larger than the angular acceptance in the observations, and relevant effects such as advection and synchrotron self-absorption have been neglected, our approximate result in Eq.~\ref{eq:fracW} puts us on the track that the strongest constraints on the WIMP parameter space should be related to synchrotron emission.

\section{benchmarks and complete treatment}
\label{sec:fulltreat}

\subsection{A few benchmark scenarios}

\begin{table}[t]
\begin{center}
\begin{tabular}{|c|c|c|c|c|c|}
\hline
$\,\,\,\,\,\,\,\,\,\,\,$&$M_{\chi}$&$\sigma v$&ann. mode&B&$\rho$
\tabularnewline
\hline
\hline
\b
B1&$100\, {\rm GeV}$&$\,2\cdot10^{-25}\, {\rm cm^3s^{-1}}$&$b-\bar b$&Equipart.&$N_{sp}$
\tabularnewline
\hline
\hline
\b
B2&$100\, {\rm GeV}$&$\,6\cdot 10^{-30} \,{\rm cm^3s^{-1}}$&$b-\bar b$&Reconnect.&$A_{sp}$
\tabularnewline
\hline
\hline
\b
B3&$1 \,{\rm TeV}$&$\,2\cdot10^{-28}\,{\rm cm^3s^{-1}}$&$b-\bar b$&Constant&$A_{sp}$
\tabularnewline
\hline
\end{tabular}
\end{center}
\caption{Benchmark models.}
\label{tabBM}
\end{table}

The Milky Way is the galaxy we know in furthest detail, still the determination of its DM halo profile is not a simple task, in particular, for what concerns its inner region. As is well known, there is some tension between N--body simulation results suggesting the presence of a sharp enhancement in the central density, with dynamical observations of spiral galaxies finding configurations consistent with a density profile with a constant central core. 

One of the uncertainties in implementing results from N--body simulations regards the interplay between dark matter and the baryonic components of the Galaxy; in particular, the formation of the supermassive black hole (SMBH) at the Galactic center could have strongly modified the initial DM profile. 
The adiabatic growth of a black hole at the center of a singular halo profile leads to the accretion of a very dense DM spike around it~\cite{Gondolo:1999ef}.
Although this picture requires tuned initial conditions \cite{Ullio:2001fb}
(SMBH seed very close to the center of the dark matter distribution and slow adiabatic growth), it is actually not excluded and, if the spike is formed, it can be completely destroyed only in a major merger event, unlikely in the recent past of the Galaxy. 
The picture sketched in \cite{Gondolo:1999ef} and  \cite{Ullio:2001fb} has been further refined in \cite{Bertone:2005hw},  where a time-evolution analysis of the cusp formation is performed, including the effects of self annihilations, scattering of dark matter particles by stars, and capture in the black hole. 

The presence of relatively large overdensity in the Galactic center region is an essential ingredient for a sizable WIMP dark matter signal at any of the wavelengths we will consider in our treatment. 
We follow the analysis  in \cite{Bertone:2005hw} and focus our attention on two distributions obtained from the evolution of a Navarro-Frenk-White (NFW) profile~\cite{Navarro:1996gj}: in the first (hereafter labeled $N_{sp}$) we include the formation of a density spike around the SMBH only, while the second profile (hereafter labeled $A_{sp}$) is obtained by taking into account the deepening in the Galactic potential well generated by the slow adiabatic formation of the stellar component in the inner Galaxy, as well as that of the SMBH. In this second case the stellar component itself leads to a steepening of the halo profile from $\rho\propto r^{-1}$ into $\rho\propto r^{-1.5}$~\cite{Klypin:2001xu}; such a steepening and ignoring a back--reaction on the DM profile tout--court stands as a limiting case among the series of results that have been obtained for the back--reaction effect in the literature, starting from different assumptions and using either analytic treatments or numerical simulations, see e.g.~\cite{Klypin:2001xu}, \cite{Prada:2004pi} and \cite{Gustafsson:2006gr}.
For both the $N_{sp}$ and $A_{sp}$ profiles  the effect of self annihilation triggers the density in the innermost region, with the final shape being fixed by the value of the ratio $(\sigma v)/M_{\chi}$~\cite{Bertone:2005hw}. There is therefore in general a non--linear dependence of the WIMP DM source, see Eq.~\ref{eqQ}, on the cross section (we have implemented such dependence in our analysis using scaling laws derived from either results given in \cite{Bertone:2005hw} or further sample cases kindly provided by the authors of  \cite{Bertone:2005hw}).

Our benchmark DM profiles are shown in Fig.~\ref{figprofiles}, together, for comparison, with the NFW profile and the "spiky" profile obtained implementing the original procedure outlined in~\cite{Gondolo:1999ef}.
Sample values for the WIMP mass and the annihilation cross section are chosen here such that the multi--wavelength constraints are not violated (verified a posteriori in section \ref{sec:result}).

We do not focus our discussion on specific WIMP models, but rather refer to a generic WIMP of given mass $M_{\chi}$ and annihilation cross section $\sigma v$ dominated by one single annihilation mode. 
If the DM annihilation into fermion is not suppressed, quarks give often the dominant branching ratio. This is the case for a gauge boson WIMP, such as the antiperiodic gauge field in~\cite{Regis:2006hc}, and for a Majorana fermion like the lightest neutralino in supersymmetric extension to the Standard Model. For this reason we choose as a benchmark annihilation mode a quark--antiquark pair, giving raise to soft spectra of secondary particles mainly through the hadronization into pions (charged or neutral) and their subsequent decay, see Fig.~\ref{fig:yield}a.

The case of a leptonic final state, such as $\tau^+ -\tau^-$, is rather different since much harder spectrum is produced. We consider the $b -\bar b$ and $\tau^+-\tau^-$ as limiting cases of a much more generic WIMP scenario.

To start our discussion on multi--frequency constraints on the GC as a WIMP DM source we first focus on three benchmark cases. Properties of the model are listed in Table~\ref{tabBM} and regard the particle physics setup as well as the dark matter profile and its reshaping by the baryonic component in the Galactic center region and the assumptions on the magnetic field profile, whose relevance is illustrated in what follows in the discussion of propagation.

\begin{figure}[t]
 \begin{minipage}[htb]{8cm}
   \centering
   \includegraphics[width=7.8cm]{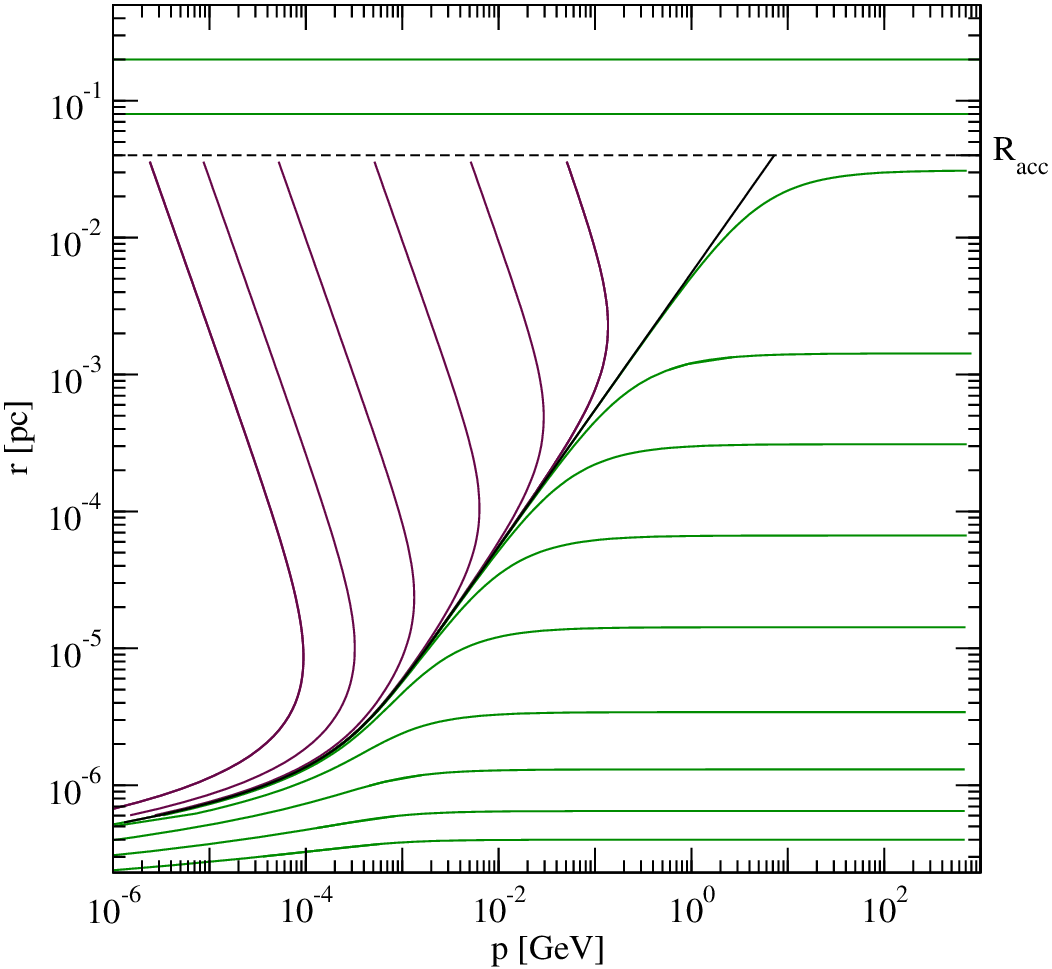}
 \end{minipage}
 \ \hspace{5mm} \
 \begin{minipage}[htb]{8cm}
   \centering
   \includegraphics[width=7.8cm]{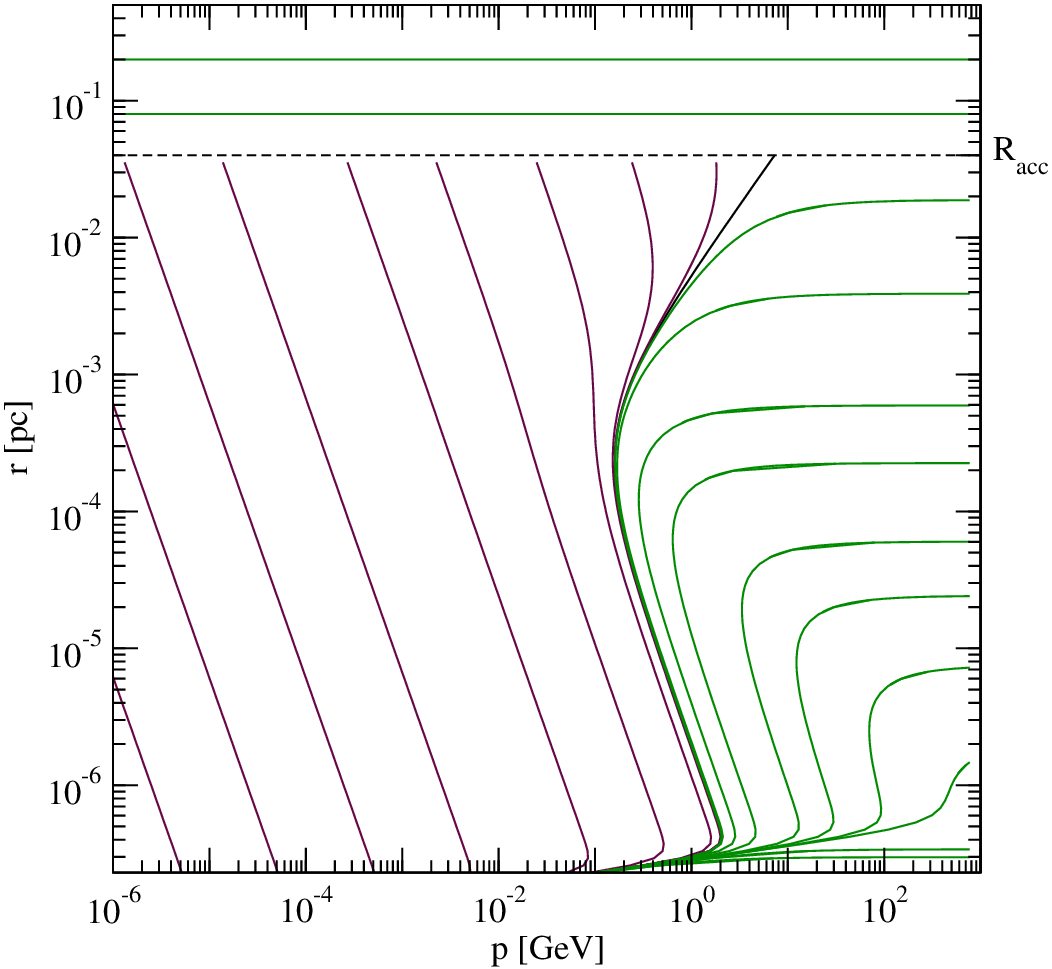}
 \end{minipage}
    \caption{Electron/positron trajectories in the plane radius versus momentum for the equipartition ({\it Left Panel}) and reconnection ({\it Right Panel}) magnetic fields. Far from the turning points, synchrotron loss is dominant in green trajectories, while adiabatic heating takes over in violet trajectories. The black solid line represents the curve along which the $e^+ -e^-$ accumulate since the two effects balance each other. The dotted line is the accretion radius $R_{acc}=0.04\,pc$, where advection is assumed to stop.}
\label{fig:traj}
\end{figure} 

To model the propagation of electrons/positrons at the Galactic center, we need to consider two regimes.
Outside the accretion flow, i.e. at radii greater than the accretion radius $R_{acc}\sim0.04$ pc, the electrons/positrons, injected by dark matter annihilations, lose energy in place through radiative processes and their equilibrium number density is simply given by Eq.~\ref{eqdnde} (we will now include all relevant radiative processes). 

For $r\leq R_{acc}$, the physical picture is as follows: The dark matter annihilations inject $e^+$ and $e^-$ at a given radius of injection $R_{inj}$;  then two competitive processes take place. On top of the momentum loss due to radiative processes, electrons and positrons gain energy in the adiabatic compression due to the plasma flow onto the central BH. The propagation equation Eq.~\ref{eqtransp} admits an integral analytic solution only in case synchrotron emission is the dominant radiative loss process and the $e^+-e^-$ are in the ultra--relativistic (or non--relativistic) regime. The solution takes the form~\cite{Aloisio:2004hy}:
{\small
\be
f(r,p)=\int_{R_{acc}}^r dR_{inj}\frac{Q(R_{inj},p_{inj}(r,p,R_{inj}))}{v(R_{inj})}\Big(\frac{R_{inj}}{r}\Big)^{4 C_{\alpha}}\Big( \frac{p_{inj}}{p}\Big)^4
\label{eqODEsolrel}
\ee   
}
where $C_{\alpha}=(2-\alpha)/3$ with $\alpha$ being the exponent in the power law scaling of the radial infall velocity  $v\propto r^{-\alpha}$, i.e. $\alpha=1/2$ in case of potential dominated by the central BH, see Eq.~\ref{eqgasvel}.
The momentum $p_{inj}$ is the initial momentum of an electron injected at $R_{inj}$, arriving at position $r$ with momentum $p$.
Outside of the ultra--relativistic approximation, Eq.~\ref{eqtransp} can be solved numerically through a change of variables that recasts the original partial differential equation (PDE) into an ordinary differential equation (ODE). This is defined by a solution of the associated homogeneous equation; the characteristic curve related to the latter is
\be
\frac{dp}{dr}=\frac{\dot p_{syn}(r,p)+\dot p_{adv}(r,p)}{v(r)} \,\,\,,\,\, p(R_{inj})=p_{inj}
\label{eqtraj}
\ee
which describes the trajectory of the electrons in the plane radius versus momentum, where
\be
\dot p_{syn}=\frac{4}{3}c\,\sigma_T\frac{B(r)^2}{8\pi} \frac{E\, p}{(m_ec^2)^2}\,\,\,\,,\,\,\,\,\,\,
\dot p_{adv}=-\frac{1}{3r^2}\frac{\partial}{\partial r}(r^2v)\,p\;.
\label{eqdotp}
\ee

The solution of Eq.~\ref{eqtraj} is shown in Fig.~\ref{fig:traj} in the plane $(p,r)$, in the case of equipartition (left panel) or reconnection (right panel) magnetic field (see Fig.~\ref{fig:Bfield}). In the first case the synchrotron loss dominates at high energies, while the advection gain takes over at low energies; electrons accumulate on the trajectory separating the two regimes (black curve in the figure). Since approaching the BH, the scaling in radius of the synchrotron loss is faster than the advection gain,  $\dot p_{syn}\propto r^{-5/2}$ versus $\dot p_{adv}\propto r^{-3/2}$, the advection dominated region becomes smaller and smaller and disappears for radii very close to the BH horizon. As stated above, in the region with $r > R_{acc}$ we neglect the advection and thus the trajectories are just horizontal lines.

Quite similar is the electron/positron flow in the case of a reconnection magnetic field. Since now the magnetic field is smaller, the advection dominated region becomes larger. The line along which electrons accumulate is modified accordingly to the shape of the magnetic field plotted in Fig.~\ref{fig:Bfield}a.

We would assume a magnetic field which is constant in the accretion region, see again Fig.~\ref{fig:Bfield}a, advection basically dominates throughout the plane and there's no region of accumulation. Moreover electrons could be accelerated at energies greater than M$_{\chi}$, something which is not possible in the previous cases where the propagation of electrons with energy $\ge\,10$ GeV becomes dominated by the synchrotron loss at all radii.

We can then solve the propagation equation Eq.~\ref{eqtransp} on these trajectories, reducing the PDE to a linear ODE that admits a standard integral solution:
{\small
\be
f(r,p)=\int_{R_{acc}}^r dR_{inj}\frac{Q(R_{inj},p_{inj}(r,p,R_{inj}))}{v(R_{inj})} \exp{\Big(\int_{R_{inj}}^r dr' \frac{h(r',p_{inj}(r,p,r'))}{v(r')}\Big)}\;,
\label{eqODEsolut}
\ee   
}
where $h(r,p)=p^{-2}\frac{\partial}{\partial p}(\dot p_{syn}(r,p)p^2)$.
In the ultra--relativistic limit Eq.~\ref{eqODEsolut} reduced to the form in Eq.~\ref{eqODEsolrel}.

\begin{figure}[t]
\begin{minipage}[htb]{8cm}
   \centering
   \includegraphics[width=7.8cm]{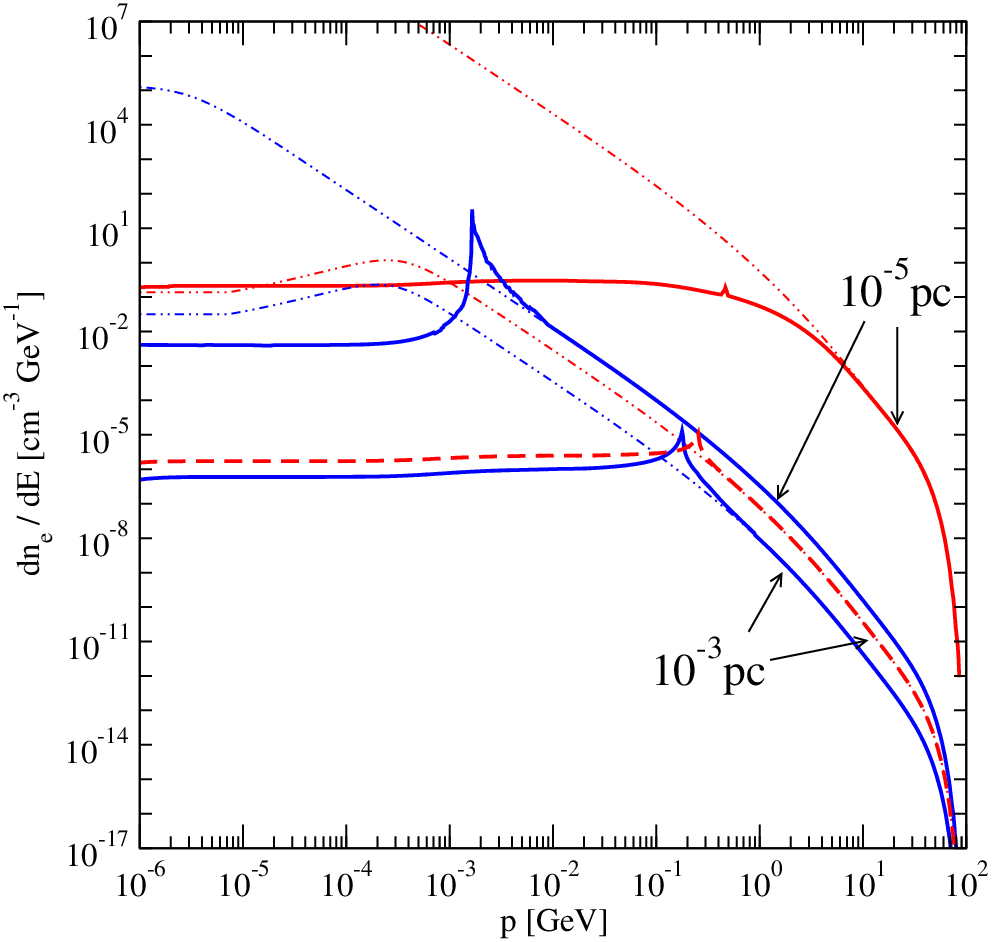}
 \end{minipage}
 \ \hspace{5mm} \
\begin{minipage}[htb]{8cm}
   \centering
   \includegraphics[width=7.8cm]{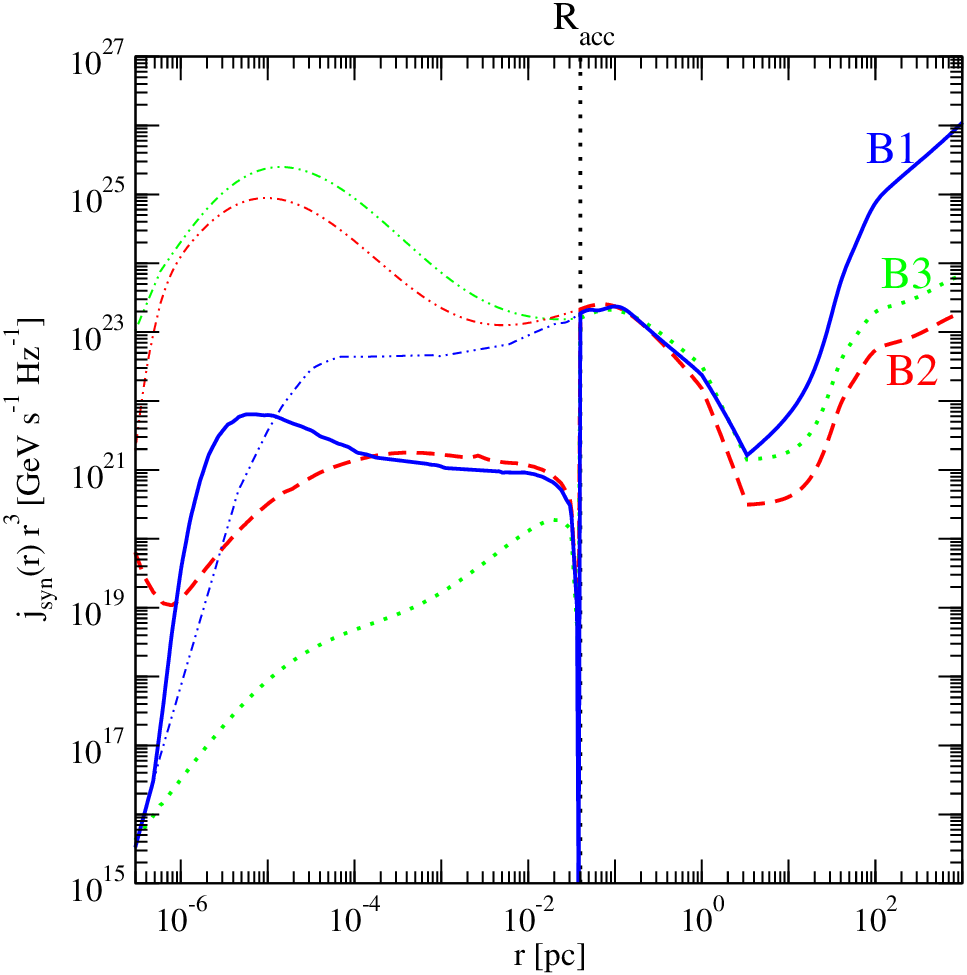}
 \end{minipage}
   \caption{{\it Left Panel:} Electron/positron equilibrium number density at two given radii for the benchmark models B1 and B2 (colors and line-styles as in the previous figures). In the dash--dotted lines the effect of advection is neglected. {\it Right Panel:} Synchrotron luminosity per unit logarithmic interval $j_{syn}\, r^3$ at 90 cm for the three benchmark models. The sharp transition is in correspondence to the accretion radius $R_{acc}=0.04\,pc$, where advection starts. In the upper dashed--dotted curves the effect of advection is neglected.}
\label{fig:adv}
\end{figure}

Examples of the resulting electron/positron equilibrium number density are plotted in Fig.~\ref{fig:adv}a.
We can see that the effect of the advection is to drive low energy electrons to higher energies, where synchrotron loss is dominant. Thus there is a peak in the distribution corresponding to the curves of momentum accumulation in Fig.~\ref{fig:traj}.
Note that in the case of equipartition magnetic field the accumulation flow is much more efficient with respect to the reconnection case, or, in other words, there is a wider region of the initial condition $(p_{inj},R_{inj})$ for a point of accumulation $(p,r)$, and thus more electrons contribute. For this reason the peak in the density are more pronounced in the equipartition case.
In Fig.~\ref{fig:adv}a we plot for comparison the electron/positron equilibrium number density obtained neglecting the effect of advection. The synchrotron losses dominate until very low energies (and not too small radii) where ionization takes over (see Fig.~\ref{fig:enloss}a) and the distribution becomes flatter.

Fig.~\ref{fig:adv}b gives a feeling for the radial reshaping of synchrotron signals due to advection effects.
We plot the synchrotron luminosity, see Eq.~\ref{eqlumin}, per unit logarithmic interval $j_{syn}\, r^3$, at the wavelength of 90 cm and for the three benchmark models in Table~\ref{tabBM}.
There is a sharp jump in the emissivity at the accretion radius $R_{acc}$ since we have assumed a sharp transition between the two propagation regimes; in a more realistic model we would find a slightly smoother behavior without, however, prediction for signals significantly affected.  
At this frequency the source is rather extended, as already pointed out with the approximate scalings in Section 3. Actually, advection reduces even further the signal from the innermost region. Indeed at large wavelengths the synchrotron power peaks at low energy, while advection shifts electrons from low to high energies. This effect is more evident for constant and reconnection magnetic fields where the region in the plane $(p,r)$ dominated by advection is large.
For shorter wavelengths, the advection effect becomes less and less important since the synchrotron power peak shifts to high energies and thus into the region of the plane $(p,r)$ in Fig.~\ref{fig:traj} where the synchrotron losses are dominant.

\subsection{Points sources or extended sources?}

\begin{figure}[t]
 \begin{minipage}[htb]{8cm}
   \centering
   \includegraphics[width=7.8cm]{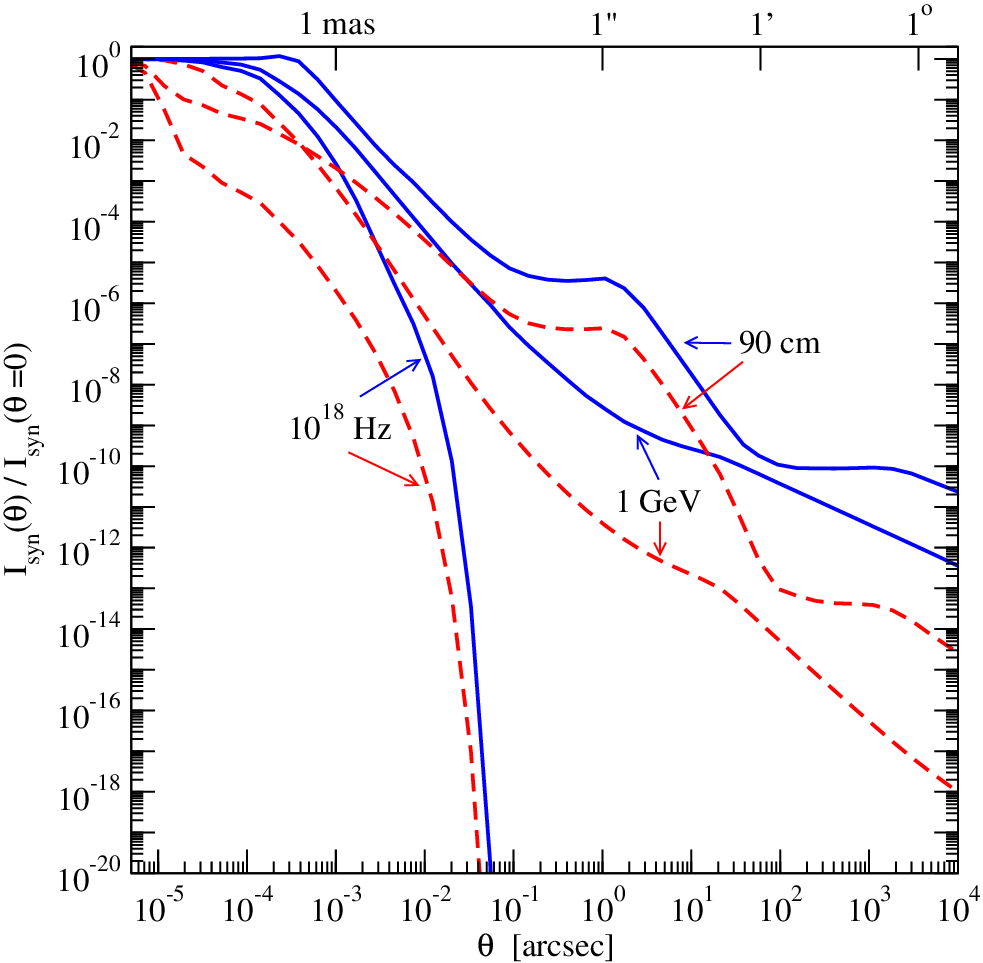}
\end{minipage}
 \ \hspace{5mm}  \
 \begin{minipage}[htb]{8cm}
   \centering
   \includegraphics[width=7.8cm]{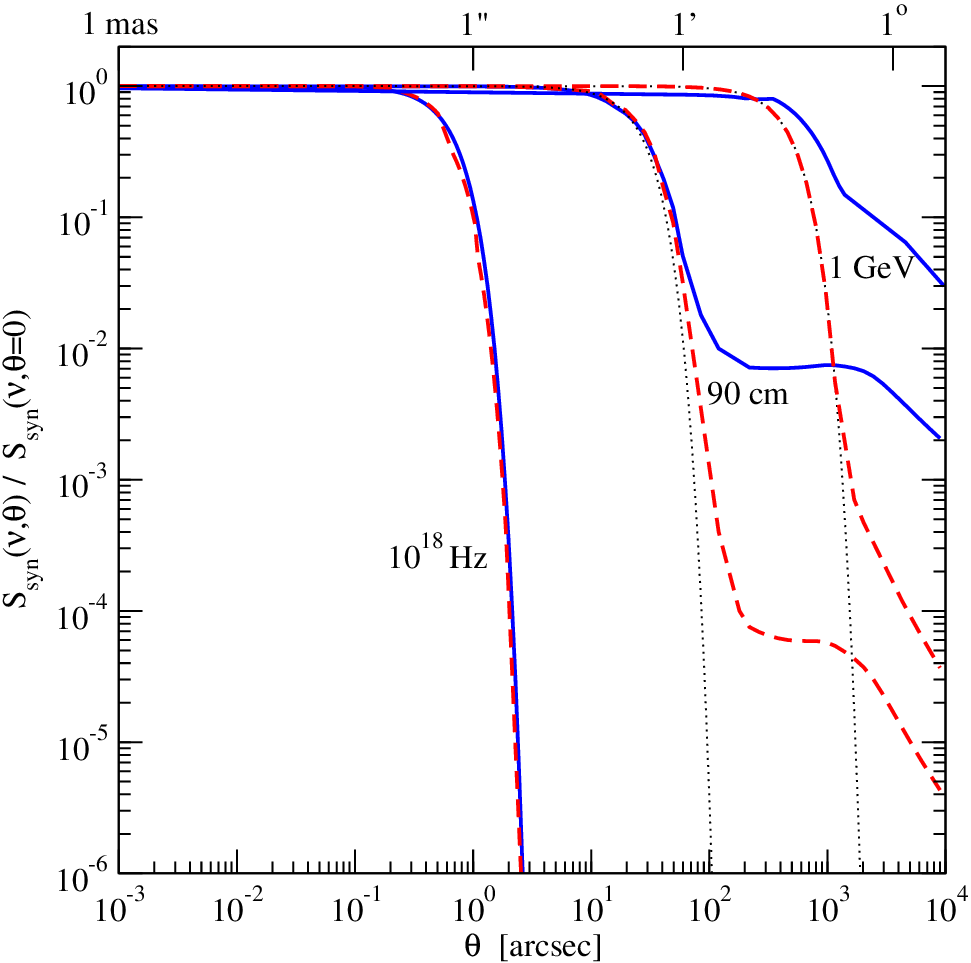}
 \end{minipage}
    \caption{Normalized radiation intensity in the radio, X and $\gamma$--ray bands for the benchmark models B1 and B2. In the {\it Left Panel}, an ideal infinite angular resolution is considered, while in the {\it Right Panel} the intensity is filtered over typical angular resolutions: $20$ arcsec at $90$ cm, $0.5$ arcsec at $10^{18}$ Hz, and $0.1^{\circ}$ at $1$ GeV. Dotted lines are the related experimental angular profiles of a point-like source, as modeled by a Gaussian detector response.}
 \label{fig:Iang}
\end{figure} 

Indirect detection of dark matter through the identification of a photon excess is not a straightforward task. There is essentially two types of signal for such flux: spectral signatures or signatures related to the morphology of the source.
Regarding the spectral signatures, prompt annihilation into monochromatic photons is the most favorable case, however it is not guaranteed in a generic WIMP model~\cite{Bergstrom:1997fj}.
On the other hand, signals with continuum energy spectrum could be in general mimicked by standard astrophysical sources. 
The spatial structure of the DM source, in case this is extended, could be an equally powerful way of  
disentangling the source from an environment in which other astrophysical sources are present.
One often has to face the problem that although the WIMP source is extended, it cannot be experimentally resolved. In the following we want to show that this is not the case at the radio frequencies, since as expected from the approximate results in Section 3 the DM source may be very extended.

To study the angular profile of the photon source induced by WIMP annihilations, we define as ideal radiation intensity $I(\theta)$ the signal in a detector with an infinite angular resolution.
For $\gamma$--rays, the spatial extension is completely fixed by the halo profile, i.e. by the dimension of the DM source. For synchrotron emission, on the other hand, it is affected by many ingredients, both related to the dark matter properties, to the magnetic field shape, and to the frequency of observation, as we can see from Eq.~\ref{eq:nuLnusg}. In case synchrotron loss is not the dominant radiative loss, also gas and starlight spatial distributions contribute to set the shape of the angular profile of the signal.
We expect from the approximate treatment the radio--band signal to become wider than that in the $\gamma$--ray band, while in the X--ray band the need of a very large magnetic field shrinks the signal to a region which is much smaller than the size of the DM source. This is confirmed in Fig.~\ref{fig:Iang}a, where we plot the intensity as a function of the angular off--set from the GC, for the benchmark models B1 and B2 in Table~\ref{tabBM} at the radio, X and $\gamma$--ray bands, normalizing each of the fluxes to unity to better understand the relative spatial extension.
The difference in the spatial extension between the two benchmark models is essentially given by the halo profiles, since the $A_{sp}$ profile leads to a more narrow signal than $N_{sp}$.

\begin{figure}[t]
   \centering
   \includegraphics[width=10 cm]{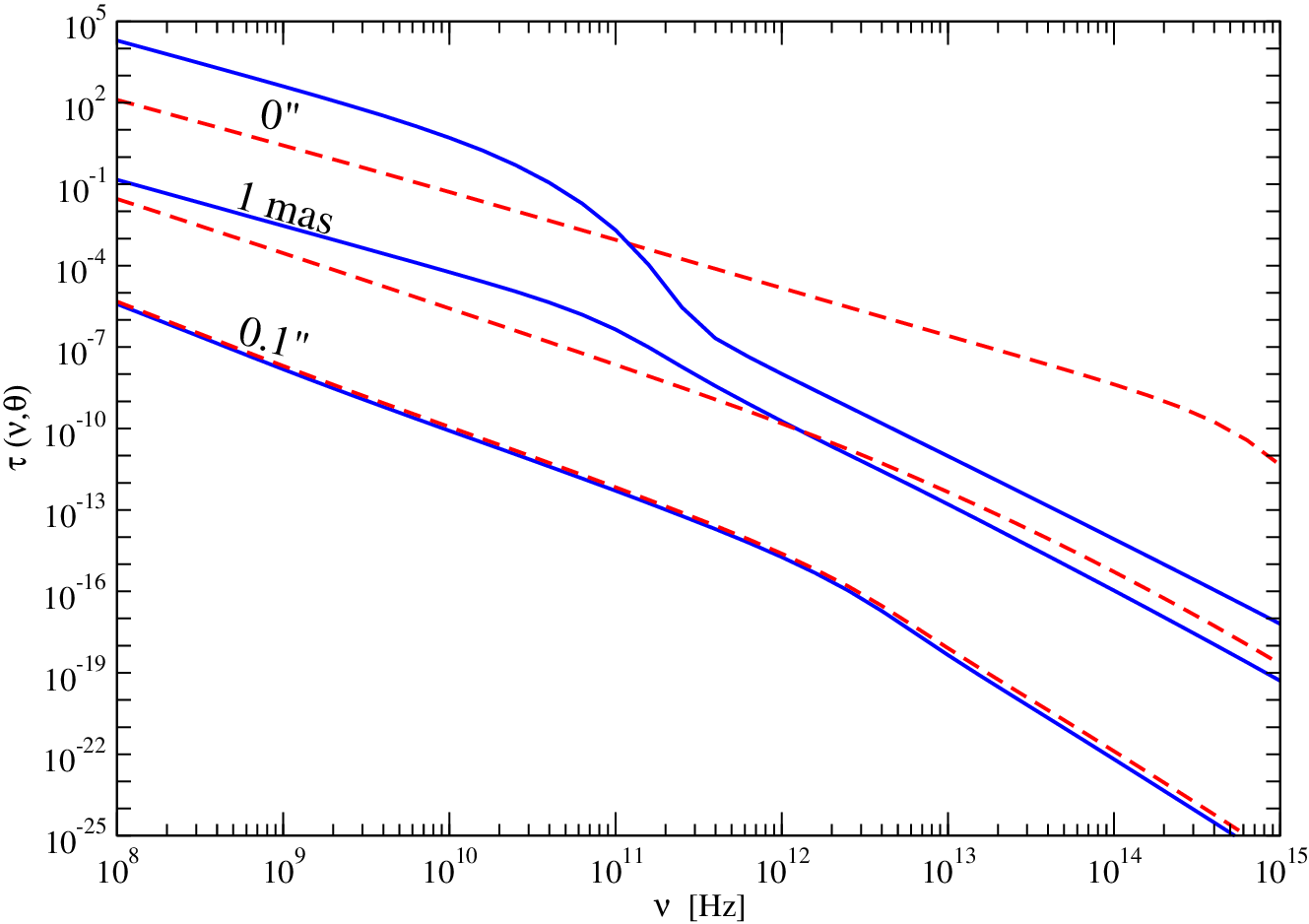}
    \caption{Optical depth versus frequency for three different lines of sight, in the benchmark models B1 and B2 (same line styles and colors of previous figures).}
\label{fig:optdepth}
 \end{figure}

In Fig.~\ref{fig:Iang} we are evaluating and including synchrotron self--absorption effects, i.e. taking into account that the emitted synchrotron radiation could be reabsorbed by the radiating electrons along the line of sight as described by Eq.~\ref{eqdiffIsynch}. Being $\alpha(\nu,s,\theta)$ the synchrotron self--absorption coefficient, see e.g.~\cite{Rybicki}, the quantity which is useful to estimate
the relevance of the absorption effect is the optical depth:
\be
\tau_{\nu}(\theta)=\int_{los}ds\,\alpha(\nu,s,\theta)\;.
\ee
In Fig.~\ref{fig:optdepth} we plot the optical depth along three different lines of sight for the benchmark models B1 and B2.
As we can see, the absorption effect is relevant only along the lines of sight pointing towards the very central region. This is due to the fact that the probability of the radiation to be reabsorbed is related to the compactness of the source.
Thus in general we expect negligible effects for shallow profiles.
The scaling of absorption with frequency, in general, takes the approximate form: $\alpha(\nu,s,\theta)\propto j(\nu,s,\theta)\,\nu^{-5/2}$ \cite{Rybicki}.
More precisely for the benchmark models, we find numerically that absorption modifies by a factor O(1) the flux associated to observations of the inner region in the radio band, while it is irrelevant at larger angles and frequencies.

In a real observation, the detected angular profile is a combination of the intrinsic profile shown in Fig.~\ref{fig:Iang}a and the experimental resolution, as described by Eq.~\ref{eqIsynch}.
In Fig.~\ref{fig:Iang}b we plot again the WIMP induced emissions for the benchmarks models B1 and B2, now filtered over a typical angular resolution. For the 90 cm signal, we take a typical resolution achievable by VLA, namely FWHM=$20''$~\cite{VLA}. For the X--rays emission we consider the Chandra point spread function, i.e. PSF=$0.5''$~\cite{Baganoff:2001ju}. Finally in the $\gamma$--ray case, the signal is integrated over $0.1^{\circ}$ that is a typical PSF for both the upcoming gamma-ray telescope in space GLAST~\cite{GLAST:2007} and the current generation of ACT \cite{Albert:2005kh,Aharonian:2004jr}.
The synchrotron emission in the X--band is very narrow and thus impossible to be resolved.
In case of a very cuspy profile, like $A_{sp}$, the source could be resolved only by radio observations, while for the $N_{sp}$ profile the source could be detected as diffuse both in the $\gamma$--ray and radio bands.

\begin{figure}[t]
 \begin{minipage}[htb]{8.5cm}
   \centering
   \includegraphics[width=8.2cm]{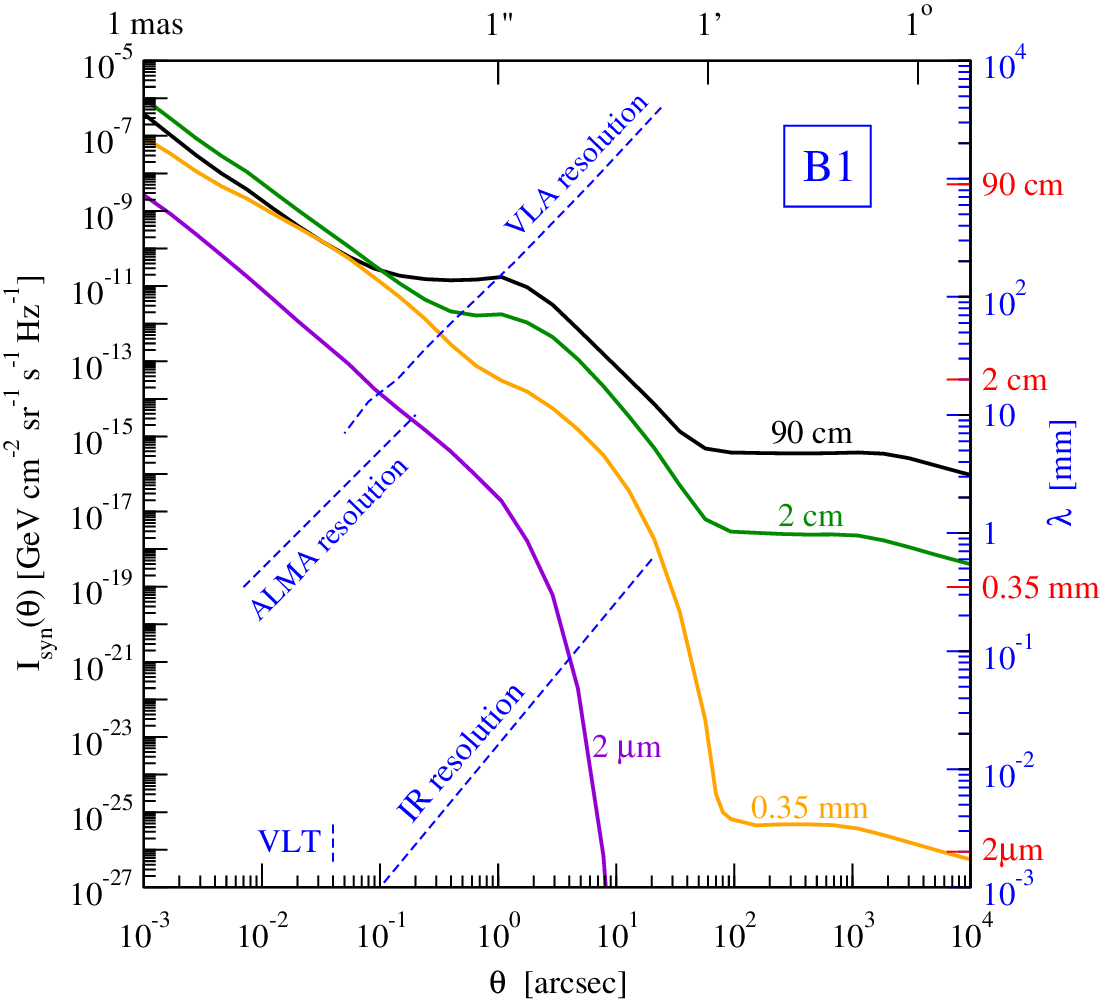}
\end{minipage}
 \  \hspace{5mm} \
 \begin{minipage}[htb]{8cm}
   \centering
   \includegraphics[width=7.8cm]{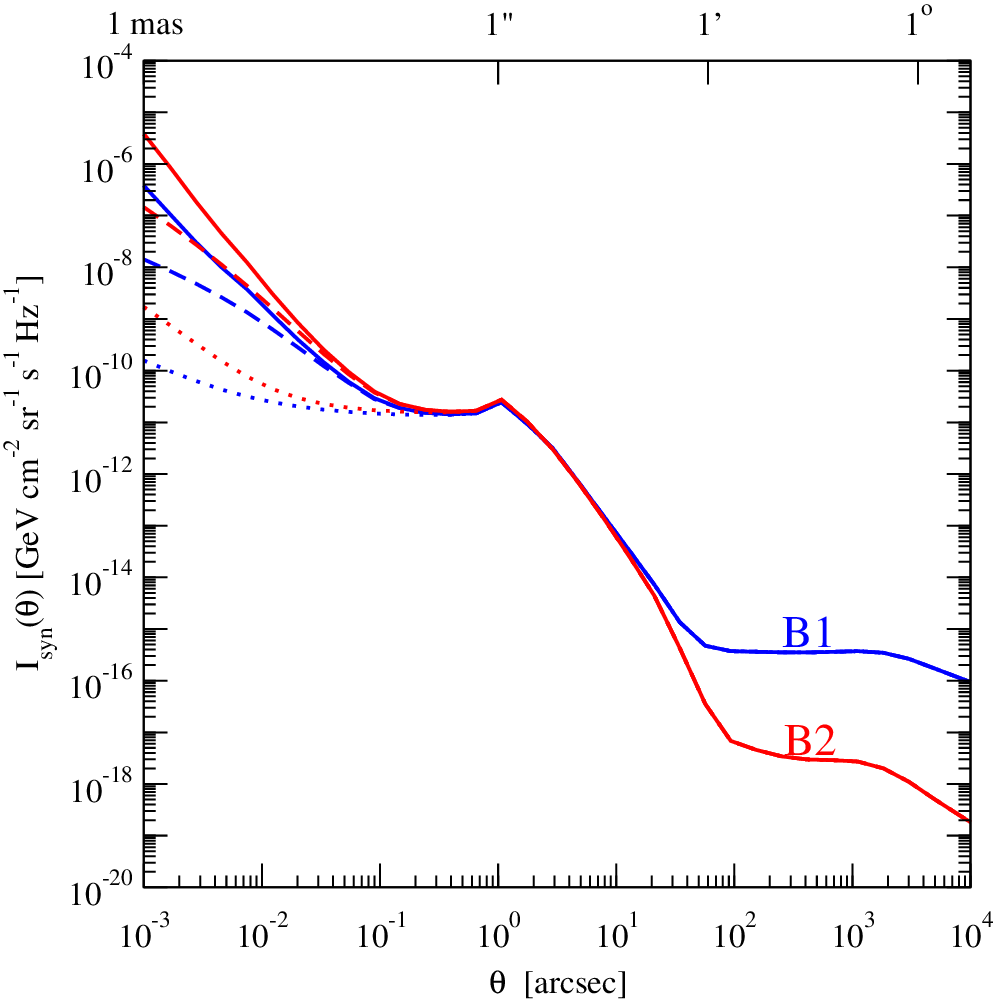}
 \end{minipage}
    \caption{{\it Left Panel:} Angular profile of the synchrotron radiation intensity at different frequencies for the benchmark model B1. Experimental angular resolution in the radio and infrared bands are also shown (blue dashed lines) with the wavelength units displayed in the scale on the right-hand side of the plot.
{\it Right Panel:} Angular profile of the synchrotron radiation intensity for the benchmark models B1 and B2 at $90$ cm, but varying the magnetic field according to the three radial profiles in Fig.\ref{fig:Bfield}a (same line styles).}
 \label{fig:InuB}
\end{figure}

In Fig.~\ref{fig:InuB}a we plot the radiation intensity for the benchmark model B1 at four different frequencies. As expected, the size of the source becomes smaller going from radio to infrared wavelengths.
In Fig.~\ref{fig:InuB}a, we show also the angular resolution of the current or near--future experiments in the radio (VLA \cite{VLA} and EVLA \cite{EVLA}), mm and sub-mm (ALMA \cite{ALMA}), infrared and NIR (VLT \cite{VLT}) bands, relative to the wavelength scale plotted on the right-hand side. 
In the first two cases the WIMP source appears extended, while going to higher frequencies it becomes a point source.

In Fig.~\ref{fig:InuB}b, we show the effect of varying the magnetic field on the synchrotron intensity at 90 cm, for the benchmark models B1 and B2, but looping over the magnetic field shapes shown in Fig.~\ref{fig:Bfield}a. Note that the choice of magnetic field differs only inside the accretion region $\theta_{acc}\sim 1''$. Equipartition choice gives the brightest signal, while the constant magnetic field the faintest. 
At this wavelength and for typical angular resolution of current detectors, the contribution from the region $\theta<1''$ is never dominant, hence the choice of the magnetic field is essentially not relevant.
The fact that at the intermediate angular scale the prediction of the two models coincides reflects just  the choice of normalizing their radio emissivity to the tightest upper bound in the radio band, i.e. $S_\nu \le 0.05$~Jy at the frequency 
$\nu=408$~MHz~\cite{davies76}.

\begin{figure}[t]
 \begin{minipage}[htb]{8cm}
   \centering
   \includegraphics[width=7.8cm]{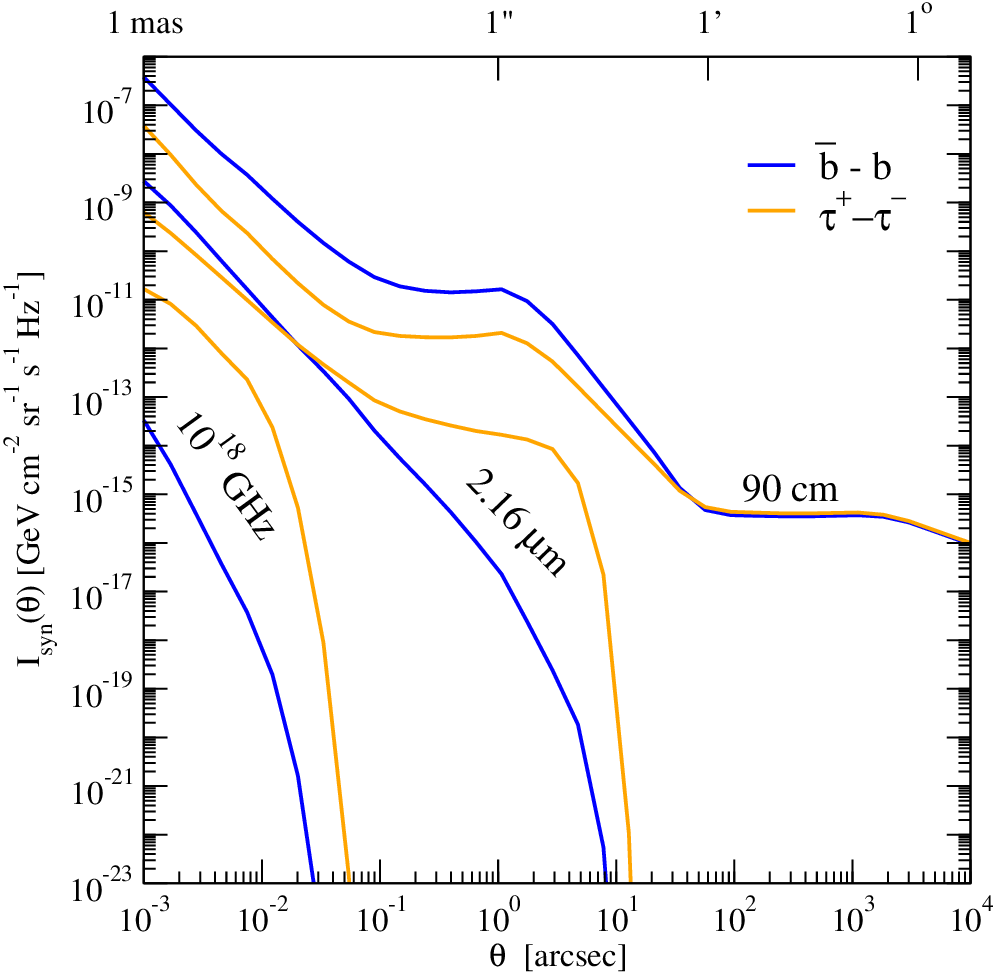}
\end{minipage}
 \ \hspace{5mm} \
 \begin{minipage}[htb]{8cm}
   \centering
   \includegraphics[width=7.8cm]{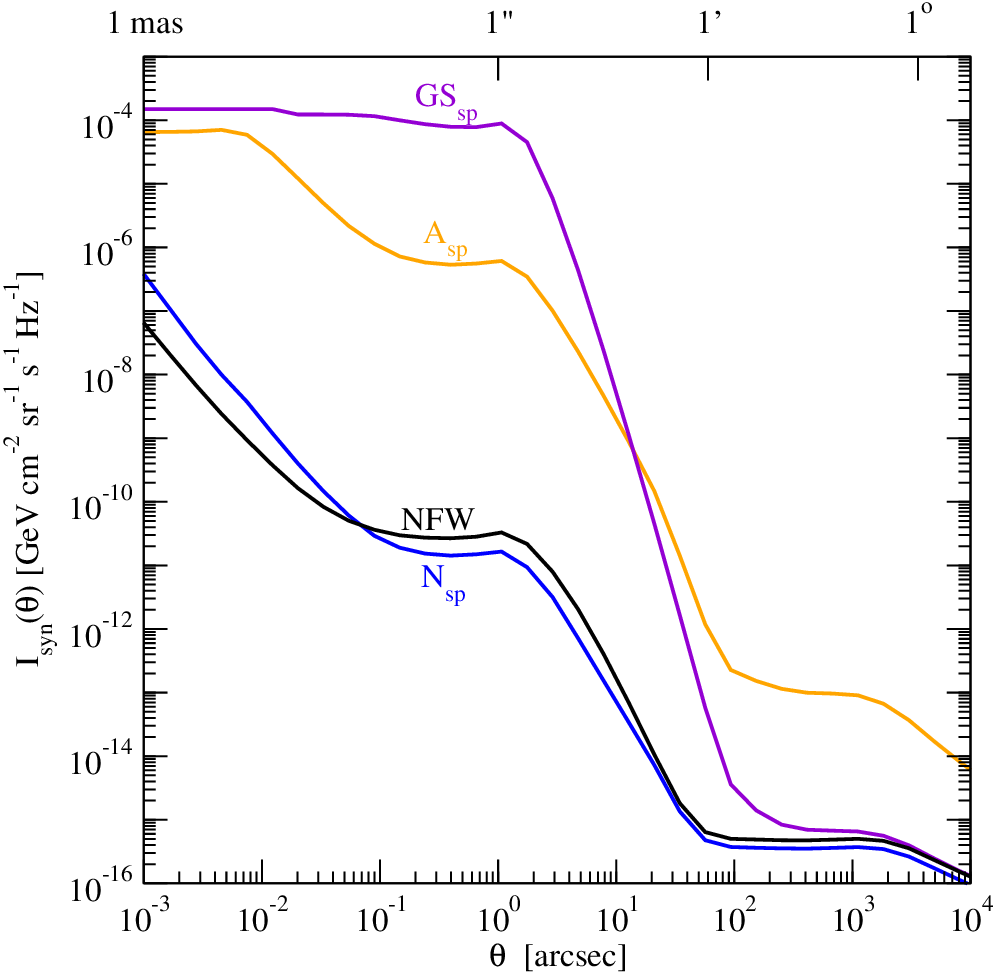}
 \end{minipage}
    \caption{{\it Left Panel}: Angular profile of the synchrotron flux intensity for the benchmark model B1 and B4 (i.e. the same of B1 except for assuming $\tau^+-\tau^-$ as dominant annihilation channel). We show the signal at different wavelengths, namely, in the radio, NIR, and X--ray bands. {\it Right Panel}: Angular profile of the synchrotron flux intensity at 90 cm for the benchmark model B1, but varying the DM halo profile.} 
 \label{fig:Ifsrho}
\end{figure}

For a given magnetic profile, the higher the energy for the radiating electron or positron, the higher the frequency at which the synchrotron power peaks; thus a hard $e^+-e^-$ spectrum emits more efficiently at short wavelengths, while a soft spectrum at long wavelengths.
In Fig.~\ref{fig:Ifsrho}a we plot the angular intensity of the synchrotron signal in the radio, NIR, and X--bands for benchmark model B1 and B4, i.e. the same of B1 except for assuming $\tau^+-\tau^-$ as dominant annihilation channel rather than $b-\bar b$. We find that B4 is significantly brighter than B1 at high frequencies. Note also that the spatial extension at such frequencies depends significantly on the WIMP annihilation final state.

Finally we turn to the uncertainty on the dimension of the signal stemming from the dimension of the source itself. 
In the case of signals at radio frequencies the scale at which is relevant to model the DM density profile to derive a definite prediction correspond basically to the angular resolution of the observation itself, i.e.  $\theta \gtrsim 1''$.
In Fig.~\ref{fig:Ifsrho}b, we plot the benchmark model B1 at 90 cm, varying the dark matter profile and find how dramatically the signal can change.
Note that the reason why the result with the NFW profile or the $N_{sp}$ profile are essentially equivalent is the large value of the ratio $\sigma v/M_{\chi}$ for the benchmark model under consideration, which is flattening out the $N_{sp}$ profile.

\section{results: multi--wavelength constraints and perspectives}
\label{sec:result}

In the previous Section, we discussed how intensity and spatial extension of the signal depend on
parameters involved in the prediction for the multi--wavelength spectra. We implement now this general analysis to derive quantitative constraints. In Section 2 we listed available data--sets on the GC relevant in our analysis; since it is unlikely that any of them comes in connection to a DM signal, we will extract upper limits only. It's not straightforward to select a uniform exclusion criterion for all the measurements. We decide to compare the DM induced flux with the most constraining data--point in any given wave--band. To some extent, this is a conservative approach, since each experimental point is considered to be independent and no correlation analysis implemented. In the next decade, new telescopes, as well as new cycles of observations with experiments already operative at present, will allow to place even tighter constraints on WIMP parameter space, or, hopefully, find evidence for a DM signal. We will focus, in particular, on two classes of $\gamma$--ray telescopes, namely, the satellite detector GLAST~\cite{GLAST:2007} which will be in orbit in a few months, and the next-generation air Cherenkov telescope CTA~\cite{CTA:2007}, and discuss the relevance of new observations at radio frequencies by the VLA project~\cite{VLA}.

\subsection{Synchrotron emission versus radio, infrared, and X-ray data}

As summarized in the Section 2, rather accurate measurements of the radio and infrared emission of the source associated to the central SMBH are available. Both the spectrum and the pattern in size of this source cannot be associated to synchrotron emission from DM annihilations. Typically, observations of Sgr~A$^*$ have been obtained with instruments with very good pointing accuracy and small angular acceptance. On the other hand, WIMP annihilations give rise to radio signals on a much larger angular size. It follows that, in general, it is incorrect to directly compare the total radio luminosity of the DM source with the luminosity extrapolated from the available Sgr~A$^*$ observations. A more accurate way of proceeding is to compute, for each model and each data--point, the DM--induced synchrotron intensity within the region corresponding to the angular resolution of the telescope, i.e. mimicking a Gaussian response of the detector with $\theta_d$ in Eq.~\ref{eqIsynch} (or a Gaussian elliptical response with two different $\theta$) as appropriate for each measurement.

\begin{figure}[t]
 \begin{minipage}[htb]{8cm}
   \centering
   \includegraphics[width=7.8cm]{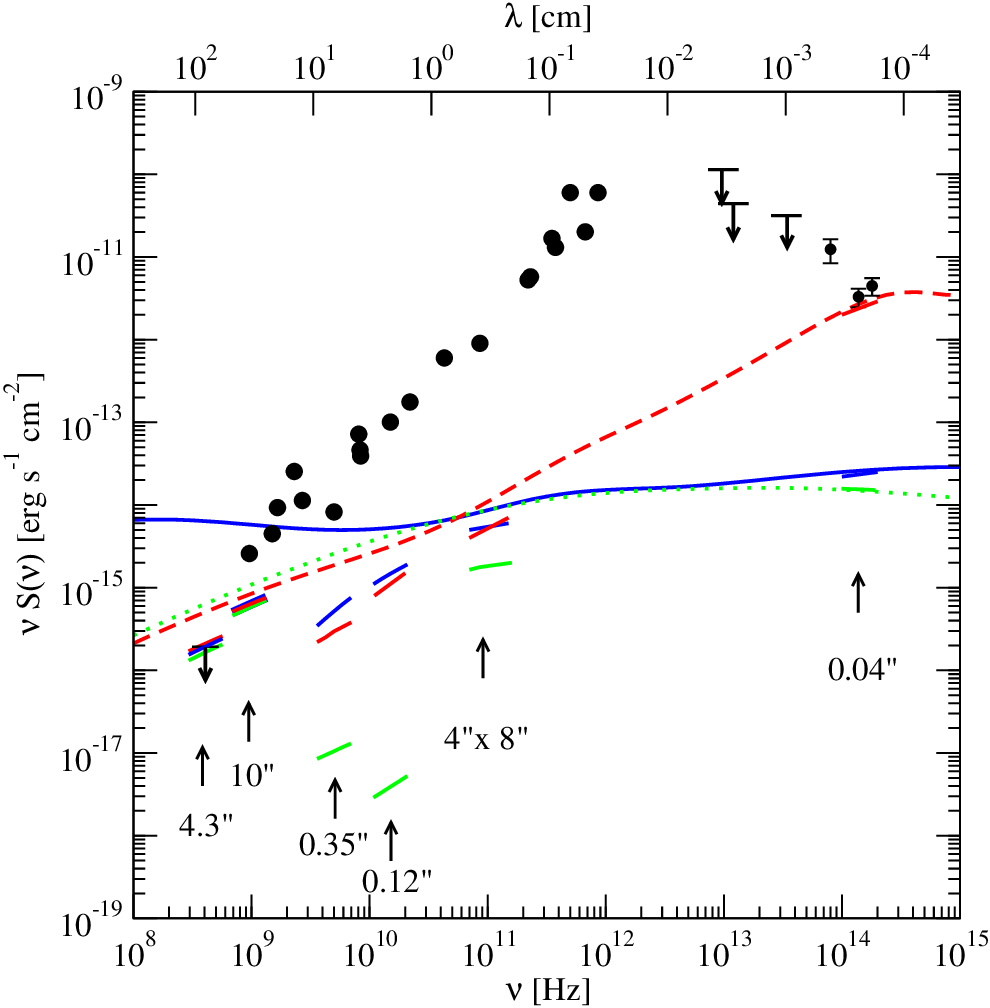}
 \end{minipage}
 \ \hspace{2mm} \hspace{3mm} \
 \begin{minipage}[htb]{8cm}
   \centering
   \includegraphics[width=7.8cm]{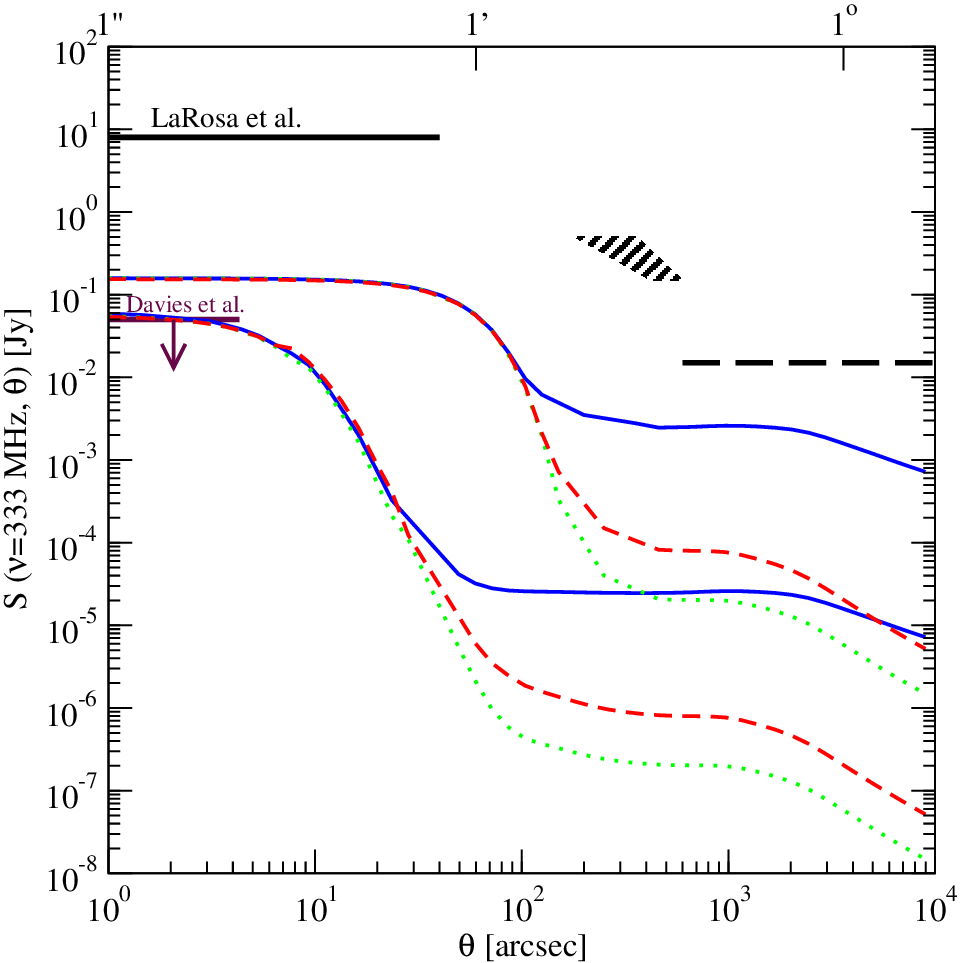}
\end{minipage}
   \caption{{\it Left Panel:} Sgr~A$^*$ luminosity in the radio and infrared bands (black points) compared to the synchrotron radiation induced by DM annihilations in the benchmark models B1, B2, and B3. For the latter, portions of spectra integrated over the experimental angular resolutions around six frequencies are shown. The lines spanning the entire range of frequencies are the spectra integrated over the whole GC region. {\it Right Panel:} Spatial profile of the DM synchrotron signal for the benchmark models B1, B2, and B3. In the upper curves the angular resolution is $43"\times 24"$, while in the lower curves it is $4.3"$. We show together the experimental limits related to the Galactic center region derived by \cite{LaRosa:2000} and to Sgr~A$^*$ by \cite{davies76}.}
\label{fig:Snu90}
\end{figure}

In Fig.~\ref{fig:Snu90}a we show measured intensities (or upper limits) for Sgr~A$^*$ (\cite{Narayan:1997ku}, \cite{Melia:2001dy}, \cite{Genzel:2003as}) together with the DM synchrotron luminosity $L(\nu)$ integrated over the whole GC region, say, e.g., a sphere of radius corresponding to an angular size of about $1^{\circ}$, and divided by $4\pi\,d_0^2$, where $d_0$ is the distance to the GC (solid, dashed, and dotted lines, respectively, for benchmark models B1, B2, and B3, spanning the whole range of frequencies shown in the plot). As we just stated, this is not the quantity which should be compared to radio data; would one make such a connection, i.e. implicitly assuming that the DM source is point-like rather than extended, the inferred upper bounds would be grossly overestimated. We select instead five data-points (plus one in the infrared), each corresponding to measurements with different angular resolutions, and plot, in a small interval around the corresponding frequency, intensities towards the GC, treating now the signal as an extended source filtered by the telescope angular response. As expected, the strongest constraint in the radio band comes from the measurement at the lowest available frequency~\cite{davies76} and the value of the cross sections for the benchmark models have been tuned to match this upper limit. This is also the measurement we will refer to, when combining constraints from different frequencies to the multi--wavelength DM spectrum in Figs.~\ref{figsvvsmb} and \ref{figsvvsmtau} below.

The intrinsic dimension of the DM synchrotron source at radio frequencies suggests that observations covering a wider field of view could set relevant constraints as well. We consider the map of the Galactic center at $\lambda$=90 cm obtained by~\cite{LaRosa:2000}, assembling different VLA observations. It is a $4^{\circ}\times 5^{\circ}$ image, with a resolution of $43"\times 24"$, thus resolving Sgr~A, the complex radio--source present at the GC and composed by Sgr~A$^*$, the supernova remnant Sgr~A East, and the spiral structure Sgr~A West, but not Sgr~A$^*$ itself. The background noise level is about 5~mJy/beam. In Fig.~\ref{fig:Snu90}b we plot the radial profile of the DM signal as it would be reported in a map with the resolution of \cite{LaRosa:2000} and detected by an observation with a resolution of $4.3"$, like in the Sgr~A$^*$ survey of \cite{davies76}. The Sgr~A source is not spherically symmetric and its angular profile cannot be accurately reduced to a radial profile; in Fig.~\ref{fig:Snu90}b we give just a schematic representation of the angular shape of the signal reported by  \cite{LaRosa:2000}.
We find that the limit on DM models one can deduce from Sgr~A data is less stringent than the constraint inferred from Sgr~A$^*$. At large angles, however, the DM signal is comparable to the background noise level, in particular, in the case of the $N_{sp}$ profile. As mentioned above, such noise level is extrapolated in \cite{LaRosa:2000}, assembling observations with different resolutions. It is not the best achievable in VLA observations today, of the order of $\sim 1\,\rm{mJy}/\sqrt{\rm{hour}}$  at 90 cm \cite{VLA}. New wide-field observations could indeed lead to tighter constraints on DM models, as we will be discuss below.

In Fig.~\ref{fig:Snu90}a we plot three measurements of the NIR luminosity of Sgr~A$^*$ in the quiescent state \cite{Genzel:2003as}, plus three upper limits derived in \cite{Melia:2001dy}, and the DM--induced signal for the three benchmark models. We discussed in some detail how the angular size of the source shrinks rapidly going to higher frequencies. For the halo profiles we consider in our analysis,  already in the NIR the DM source would appear as point--like, even with a detector with excellent angular resolution, such as a size of tens of mas achievable by VLT~\cite{Lenzen:1998,Rousset:2000}. 
Indeed, one can see that the estimate of the signal computing $L(\nu) / (4\pi\,d_0^2)$, or $S(\nu)$ 
integrated over the appropriate angular size, essentially coincide. Measurements are not far above from the estimated DM luminosities, especially for the benchmark model B2, for which this limit is comparable to the radio limit. We will derive limits on WIMP masses and cross sections considering the tightest NIR limit, namely, the measured emission in the $K_{s}$ band ($2.16\,\mu m$).

\begin{figure}[t]
 \begin{minipage}[htb]{7.5cm}
   \centering
   \includegraphics[width=7.3cm]{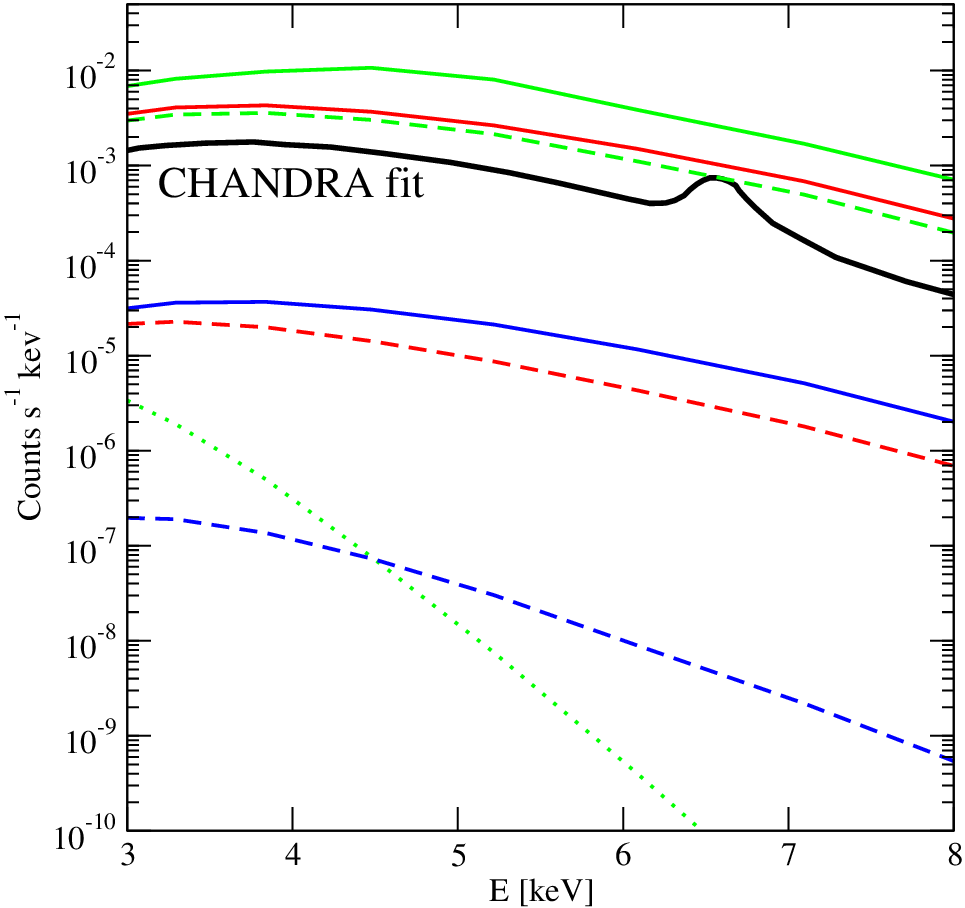}
\end{minipage}
 \ \hspace{2mm} \hspace{3mm} \
 \begin{minipage}[htb]{8cm}
   \centering
   \includegraphics[width=7.8cm]{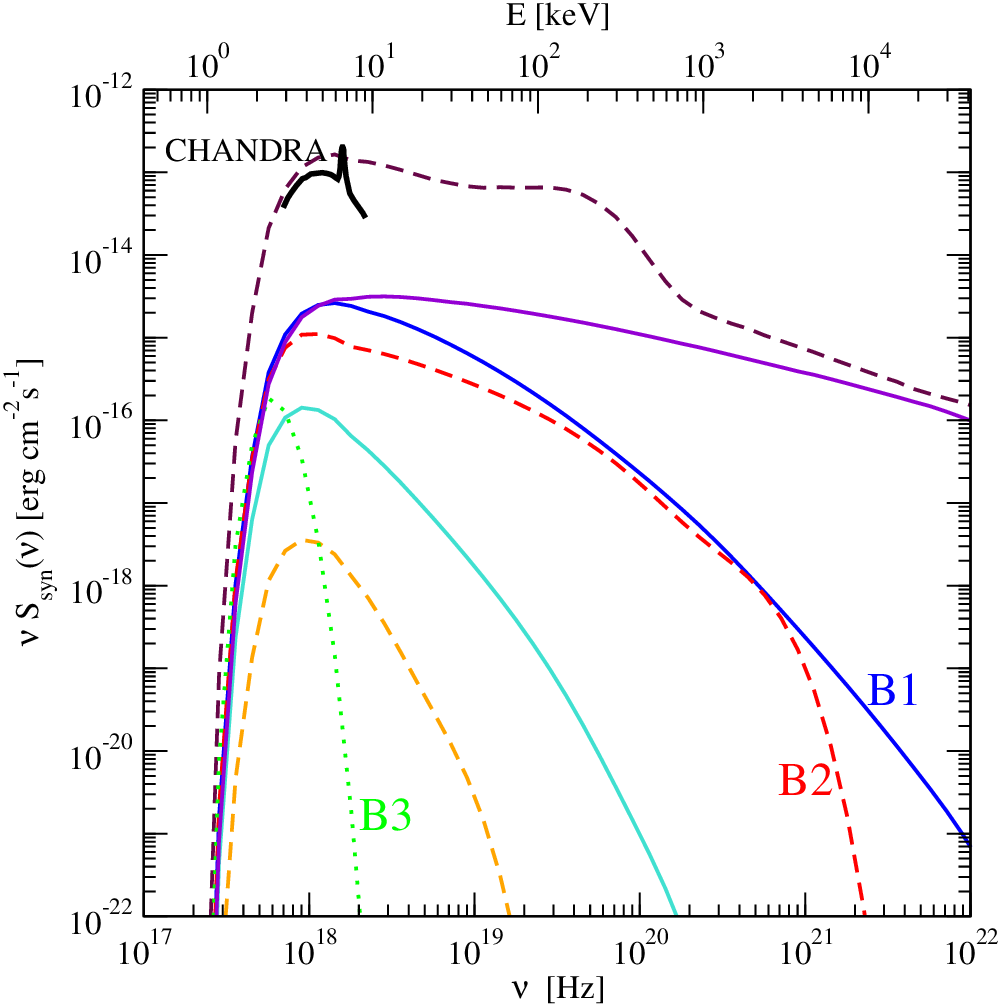}
 \end{minipage}
    \caption{{\it Left Panel:} DM induced synchrotron flux in the Chandra energy range for the benchmark models B1, B2 and B3, but varying the magnetic field among the three different shapes of Fig.~\ref{fig:Bfield}a. The black line is the fit to the Chandra measured spectrum. {\it Right Panel:} Synchrotron X--ray spectrum originated from dark matter annihilations in the benchmark models B1, B2, and B3, but varying the mass. The three cases considered are (from top to bottom): $1$ TeV, $100$ GeV, and $10$ GeV. For constant magnetic field only the first case is shown since smaller masses cannot give a sizable spectrum.}
 \label{fig:SnuX}
\end{figure}

Significant synchrotron emission at even higher frequencies is expected in the case of very large magnetic fields close to the central black hole, as in the equipartition and reconnection magnetic field models we are considering. For the flux emitted in the UV and soft--X band, we need to take into account the attenuation due to the photoelectric effect on the interstellar dust. We model this effect scaling down the emissivity of Eq.~\ref{eqjsynch} by the factor $exp(-N_H \,\sigma_{p.e.})$, where $N_H$ is the electron column density \cite{Xu:2005et} and $\sigma_{p.e.}$ is the photoelectric cross section \cite{Morrison:1983hg}.
In Fig.~\ref{fig:SnuX}a, we plot the DM signal due to synchrotron emission, in the energy range where 
Chandra \cite{Baganoff:2001ju} has detected an X--ray source with position consistent with Sgr~A$^*$.
The three benchmark models are considered, as well as the cases in which, keeping all the other parameters in the model fixed, the other choices for the magnetic field radial profile (see Fig.~\ref{fig:Bfield}a) are implemented. To convert flux intensities into counts per unit energy and time, we use the Chandra effective area on axis reported in \cite{Zhao:2004}. For a WIMP with mass of about $1$~TeV (upper green curves) the peak in the emissivity is at galactocentric distances at which equipartition and reconnection magnetic fields differ only slightly, and thus the relative signals do not differ dramatically. In the case of the  magnetic field flattened to a constant value (dotted green curve), on the other hand, synchrotron emissivity is sharply suppressed. For $100$ GeV WIMPs  (blue and red curves), the signal originates in a much smaller region, where equipartition and reconnection magnetic fields differ substantially, and the constant magnetic field cannot give a sizable signal.
To better understand the dependence on the WIMP mass of the synchrotron signal, we show the X--rays spectrum in Fig.~\ref{fig:SnuX}b for the benchmark models, and consider three WIMP mass scales.

\subsection{Inverse Compton scattering and the emission in the X-ray and $\gamma$-ray bands}

At X-ray frequencies and above, the dominant radiative process involving the $e^+-e^-$ produced by WIMP annihilations can be inverse Compton scattering, rather than synchrotron emission. IC  on the cosmic microwave background is peaked in the X--band, while that on the starlight has its peak in the multi~MeV or even GeV region. The distribution of starlight in the Galaxy is highly non--uniform; its average energy density in the inner region is about $\epsilon_*\simeq 8$ eV cm$^{-3}$~\cite{Porter:2008ve}. As a sample ansatz 
to make an estimate of the level of IC emission on starlight, we assume that such a value can be representative for the whole GC region and for simplicity we will make also the approximation of the starlight spectrum black--body shape of temperature $T_*=0.3$ eV~\cite{Porter:2008ve}.

\begin{figure}[t]
   \centering
   \includegraphics[width=10 cm]{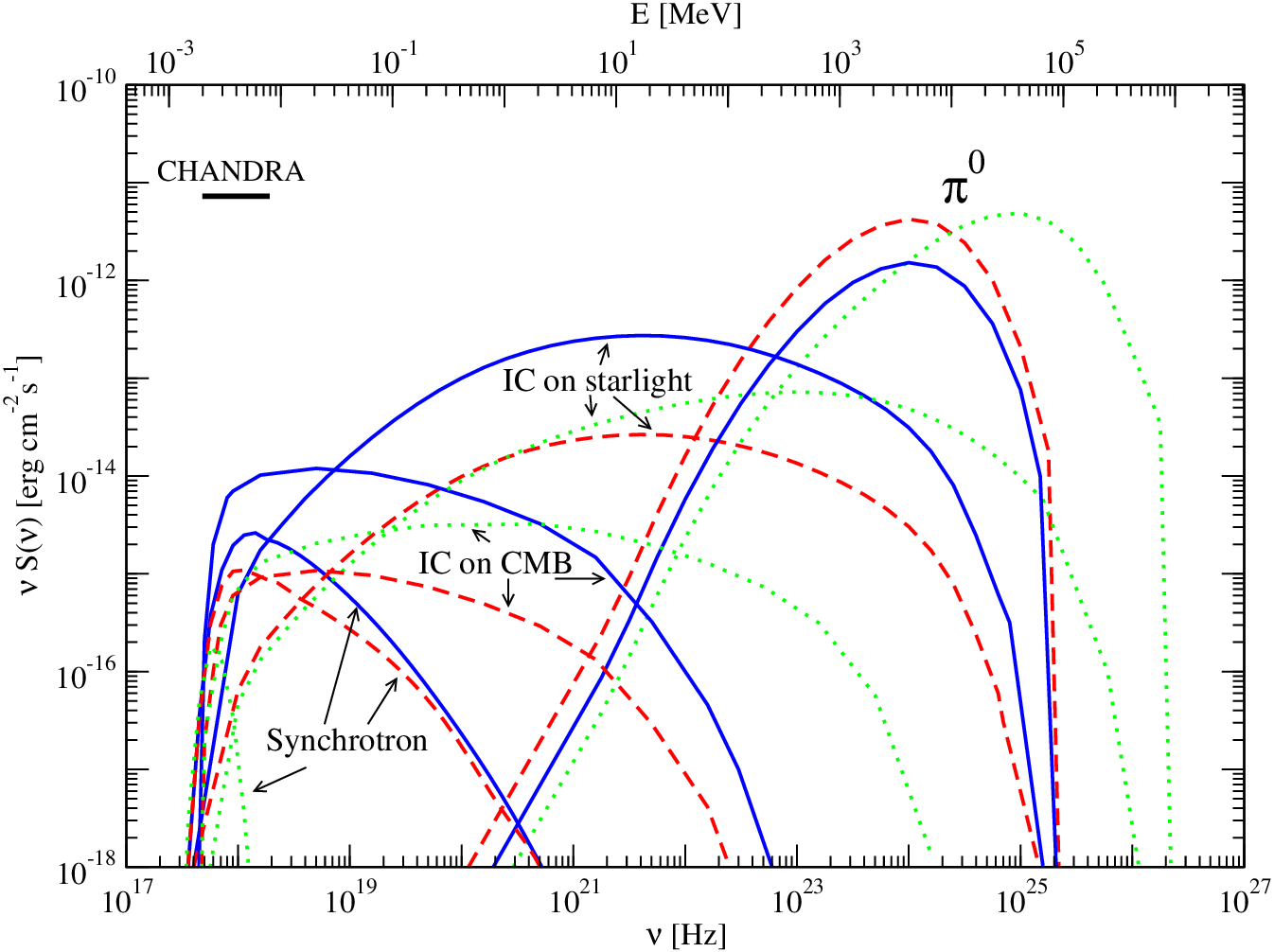}
    \caption{X--ray to $\gamma$--ray emissions induced by DM annihilations for the benchmark models B1, B2, and B3. All the four mechanisms of photon spectrum production considered in the paper give sizable signals. The flux intensities are integrated over a solid angle of $10^{-5}$ sr. The level of the diffuse emission detected by Chandra is also shown (black line).}
\label{fig:Sicgam}
 \end{figure}

In Fig.~\ref{fig:Sicgam}, we plot the IC spectra on CMB and starlight, induced by WIMP--annihilations in the three benchmarks models. It is shown for a typical angular resolution of the current $\gamma$--rays experiments, i.e. $10^{-5}$ sr. We are considering such a large field of view since the IC signals have an angular shape which is significantly broader than the shape of the $e^+-e^-$ source function. We can intuitively understand this feature from the fact that this emission comes mostly in connection to the $e^+-e^-$ with largest energy at emission, and these in turn lose energy by synchrotron losses much more efficiently close to the GC, where magnetic fields are the largest, than in the outskirts of the GC region. 
It turns out that the angular shape for the equilibrium number density of high energy $e^+-e^-$ is much broader than the gamma-ray flux from $\pi^0$ decays (which is the same as for the source function), and, of course, even more with respect to the shape of the synchrotron induced X-ray flux.  
For this reason, although for the plot in Fig.~\ref{fig:Sicgam} the intensity associated to the IC on CMB is larger than the synchrotron intensity, when integrating over the angular resolution of the Chandra detector, the trend is reversed, and only in the case of constant magnetic field, with synchrotron emission in the X-ray band essentially negligible,  comparing the IC flux to Sgr~A$^*$ gives a tighter constraint. 
Analogously to what we did in the case of radio emission, it is worth checking whether data on a large field of view could be relevant. We compare the IC signal to the diffuse X--rays emission detected by the Chandra observatory: In the $17'\times17'$ map of  \cite{Muno:2004bs}, some regions are selected and from them spectra of diffuse emission are extracted, removing events near points source and filamentary features. When combining constraints from different frequencies in Figs.~\ref{figsvvsmb} and \ref{figsvvsmtau} below, we compute the level of IC emission in such regions and extract upper bounds.

Similar arguments apply for the IC on starlight and the $\gamma$--ray limits. Indeed for what concerns bounds associated to the point--like source detected by Egret at the GC (actually its position is controversial, see the next section), the  limit associated to $\pi^0$ decay is more constraining than the IC limit. This is not true in general for the diffuse emission on the whole GC region, however we do not find any region in the parameter space in which tighter limits come in connection to this component. Note that the assumption we made on radial profile and energy spectrum for the starlight background are rather crude, and may deserve further study; refining them may lead to a slightly different conclusion, but it is unlikely that the general picture would be affected.

\subsection{The emission from $\pi^0$ decays and the $\gamma$-ray band}

Recently, observations by atmospheric Cherenkov telescopes detected a gamma-ray source in the 
direction of the Galactic center. In particular the H.E.S.S. collaboration (\cite{Aharonian:2004wa}, \cite{Aharonian:2006wh}) has obtained an accurate measurement of the spectrum of the source as a single power law in the energy range between 160~GeV and a few tens of~TeV, making the interpretation of the signal in terms of WIMP DM pair annihilations rather unlikely. H.E.S.S. has found evidence for a GC point--like source, namely, a source with an extension smaller than its PSF=$0.1^{\circ}$ and position compatible with Sgr~A$^*$, on top a diffuse $\gamma$-ray component~\cite{Aharonian:2006au}. In the case of cuspy dark matter halo profiles,  one needs to compare against the central source only; the shallower the profile, the more efficient it becomes to extend the analysis and include the GC ridge as well (see, e.g., the discussion in~\cite{Dodelson:2007gd}). The resulting limits for the benchmark profiles are plotted in Figs.~\ref{figsvvsmb} and \ref{figsvvsmtau}.

The EGRET telescope mapped the GC in the energy range 30~MeV--10~GeV~\cite{MayerHasselwander:1998hg}, detecting a flux within $1.5^{\circ}$ of the GC. A few hypothesis for interpreting this flux in terms of a standard astrophysical source have been formulated; its spectral shape is even compatible with a component from WIMP DM annihilations~\cite{Cesarini:2003nr}. On the other hand, the poor angular resolution of EGRET does not allow for a univocal identification of the source. In Ref.~\cite{Hooper:2002ru}, using only energy bins above $1$ GeV and a spatially unbinned maximum likelihood analysis, the authors argue that the Galactic center is excluded as the position of the source at $99.9\%$ and the maximum likelihood location is at $l=0.19$, $b=-0.08$. Thus they derive upper limits on the $\gamma$--rays flux from DM annihilations under the condition of no evidence of a point--source at the GC. Whether this is the correct approach is still under debate and only GLAST will give a definitive answer. We derive more conservative but robust limits comparing with the EGRET source; would one follow the line of~\cite{Hooper:2002ru}, the limits would be improved up to  about a factor of ten. Except for very light WIMPs, the strongest constraint comes from the last data-point in the EGRET measurement, in the energy bin $4-10$~GeV.

\subsection{Combined constraints on the WIMP parameter space}
 
Having specified how individual constraints are implemented, we are now ready to discuss the global picture. We refer to a model independent scenario in which a WIMP model is labeled by the value of the WIMP mass $M_\chi$ and its total annihilation rate $\sigma v$, both assumed as free and independent parameters. As for the benchmark cases, we discuss as extreme cases for the WIMP source functions, a soft spectrum configuration fixing to 1 the branching ratio in the $b -\bar b$ channel, and a hard spectrum setup when $\tau^+-\tau^-$ is the dominant annihilation mode. 
Again, having specified the annihilation mode and the WIMP mass, injection spectra are fixed accordingly to simulation results with the PYTHIA package as implemented in \ds~\cite{Gondolo:2004sc}. Reference models for the DM distribution in the GC region are the $N_{sp}$ and $A_{sp}$ profiles (with the second much denser than the first, hence with upper bounds on $\sigma v$ expected to shift dramatically). Finally, we loop over the three reference magnetic field radial profiles given in Fig.~\ref{fig:Bfield}a.

\begin{figure}[t]
 \begin{minipage}[htb]{8cm}
   \centering
   \includegraphics[width=7.8cm]{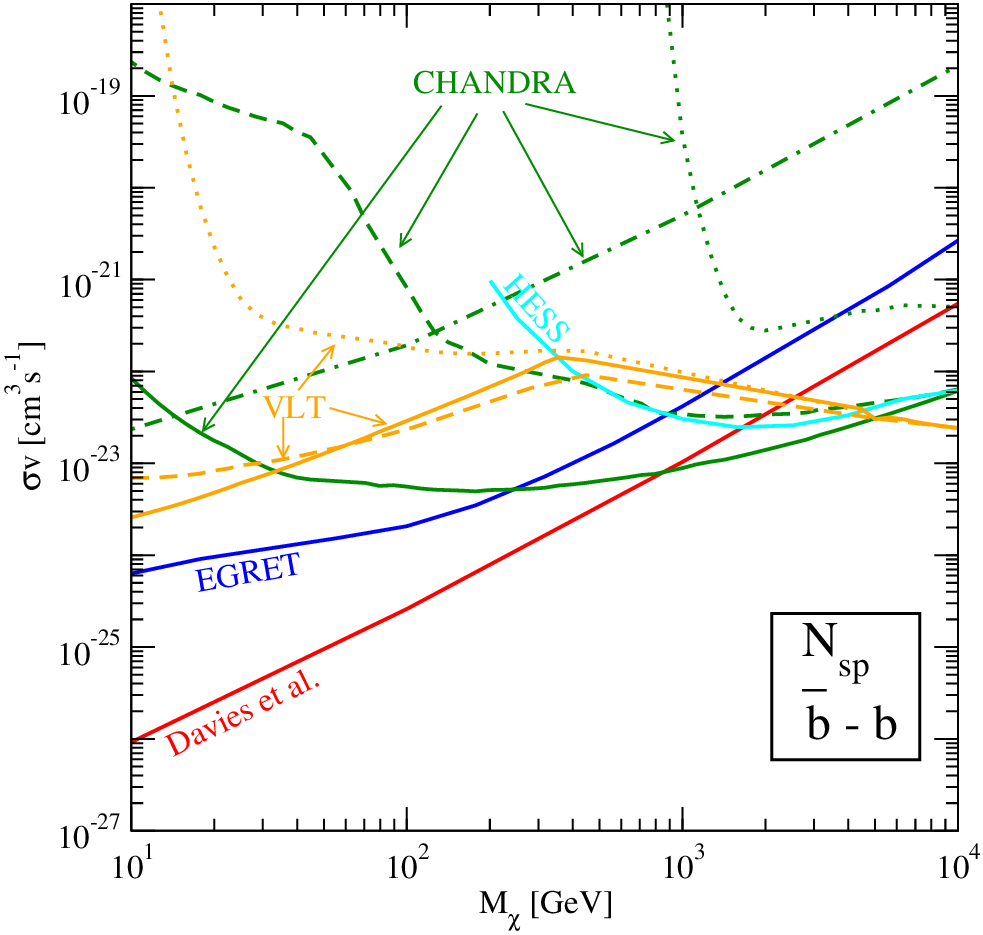}
\end{minipage}
 \ \hspace{2mm} \hspace{3mm} \
 \begin{minipage}[htb]{8cm}
   \centering
   \includegraphics[width=7.8cm]{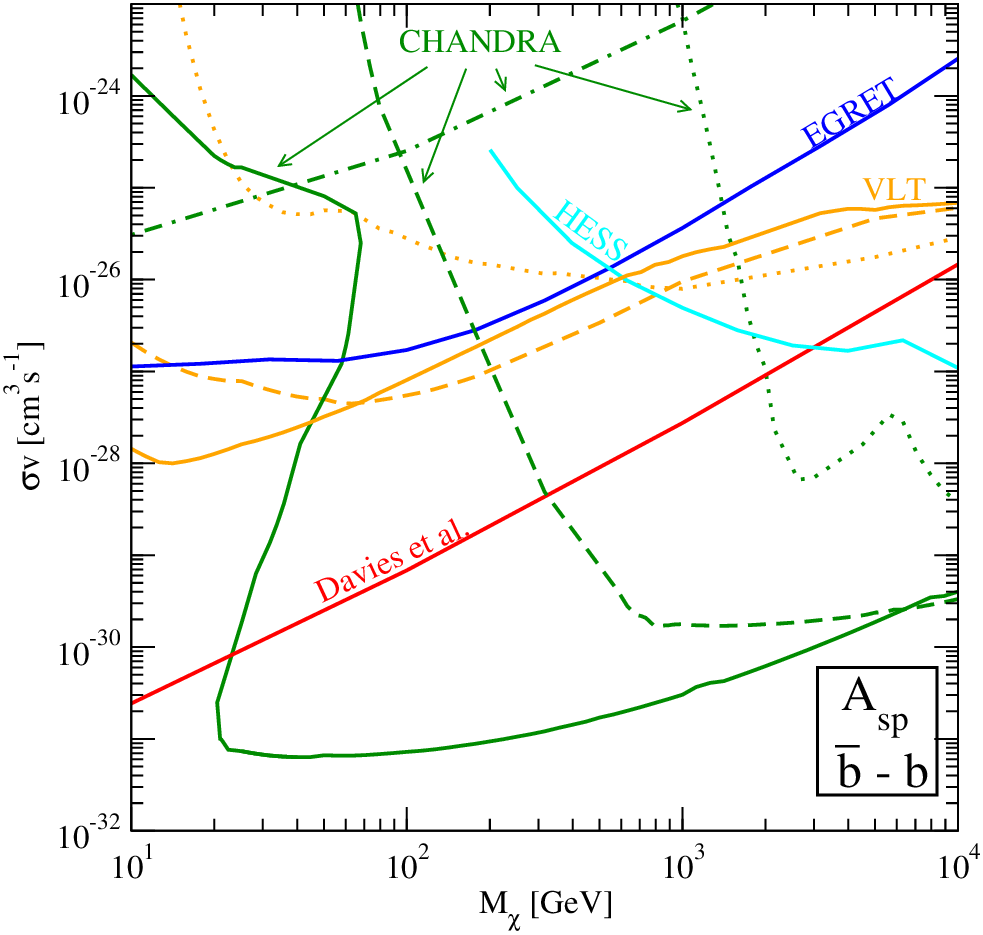}
 \end{minipage}
 \caption{Upper bounds on the WIMP pair annihilation cross section as a function of the WIMP mass, assuming $b-\bar b$ as dominant annihilation channel. The {\it Left Panel} and {\it Right Panel} show the limits for, respectively, the $N_{sp}$ and $A_{sp}$ profiles; note the mismatch on the vertical scale in the two plots. The radio constraints from Davies et al., the limits from EGRET and HESS $\gamma$-ray measurements, and the bound from the X-ray diffuse emission as detected by CHANDRA (dashed--dotted line), do not depend on the choice of the magnetic field radial density profile. The constraints associated to the NIR and X-ray observations of Sgr~A$^*$, respectively, by VLT and CHANDRA,
are shown for the three magnetic field models of Fig.~\ref{fig:Bfield}a  (using the same line styles). }
\label{figsvvsmb}
 \end{figure}

\begin{figure}[t]
 \begin{minipage}[htb]{8cm}
   \centering
   \includegraphics[width=7.8cm]{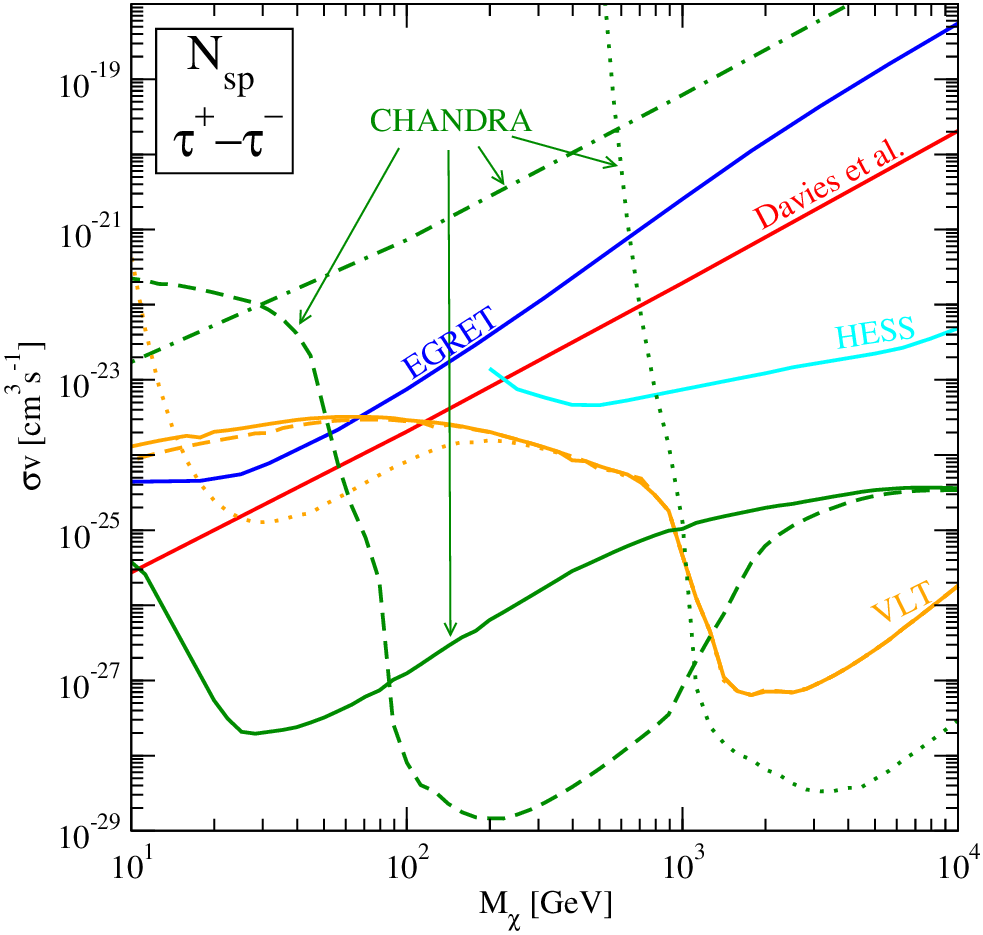}
\end{minipage}
 \ \hspace{2mm} \hspace{3mm} \
 \begin{minipage}[htb]{8cm}
   \centering
   \includegraphics[width=7.8cm]{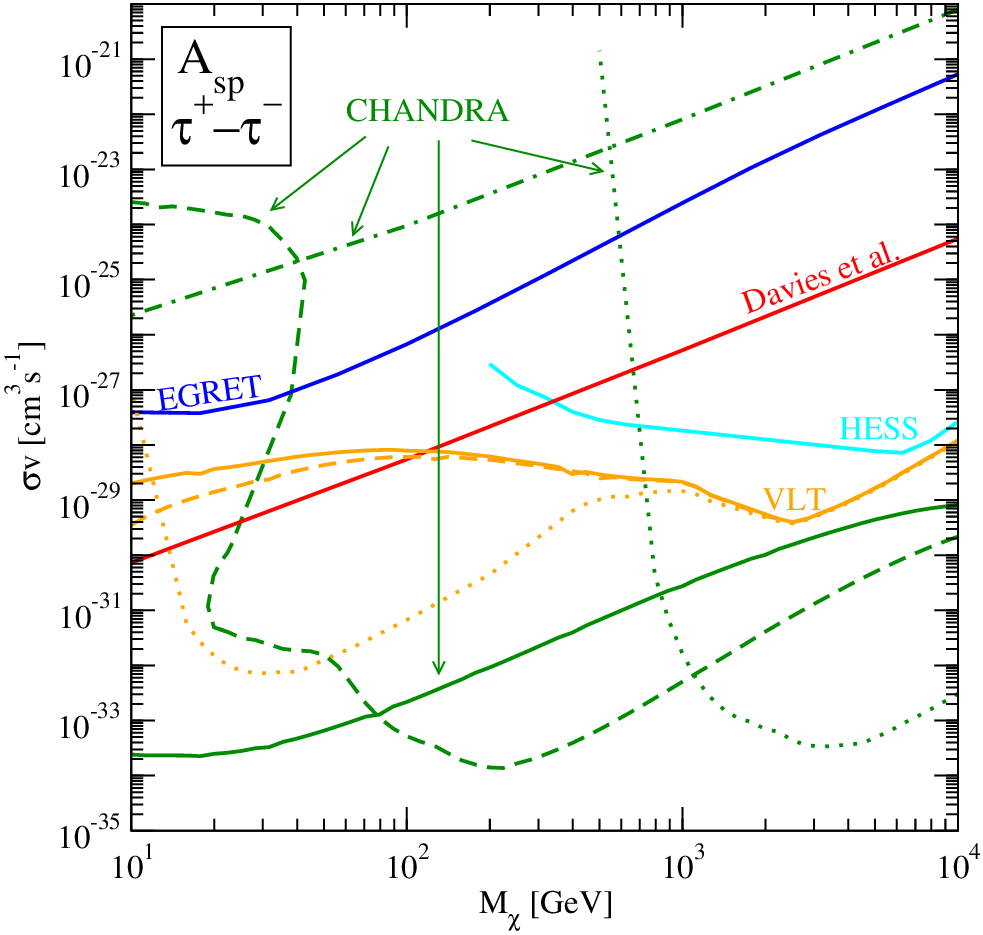}
\end{minipage}
  \caption{The same as Fig.~\ref{figsvvsmb}, but taking $\tau^+-\tau^-$ as the dominant annihilation channel.}
 \label{figsvvsmtau}
 \end{figure}

In Figs.~\ref{figsvvsmb} and \ref{figsvvsmtau} we consider the four possible combinations of annihilation channels and halo profile. The Davies et al. radio bound does not depend on the magnetic field choice since, as we have seen above, the signal is generated mainly outside the accretion region. The same is of course true for the EGRET and HESS $\gamma$-ray limits. It is rather striking to see that the radio limit is always tighter than the EGRET limit, with this trend  getting enforced even more, the softer the spectra and the more cuspy the halo profile. Were we considering a DM profile obtained by implementing the original simplified procedure by Gondolo and Silk as response for the adiabatic formation of the central SMBH~\cite{Gondolo:1999ef}, we would find that essentially the whole WIMP parameter space is excluded, as in the original conclusions in Refs.~\cite{Gondolo:2000pn,Bertone:2001jv} (despite the fact that several ingredients in this analysis are refined and/or treated differently). The HESS limit becomes more stringent for heavy WIMPs, especially in the case of hard emission spectra. Unfortunately this is a regime in which other constraints  take over.

VLT NIR limits depend to some extent to the magnetic field choice and show some non--trivial behavior. Consider the case of the $\tau^+-\tau^-$ final state. For very heavy WIMPs, and hence very energetic radiating particles in this hard emission spectra, the value of the magnetic field matching the peak in synchrotron emission is quite small (see Fig.~\ref{fig:Bfield}b),  corresponding to the region where we have assumed identical shapes for the profile of the three benchmark cases. Going to smaller masses, the energy at which the $e^+-e^-$ distribution peaks becomes smaller, and thus the required magnetic field higher, approaching the value we assigned (by mere chance) to the central plateau in the constant magnetic field case (limits are coded in magnetic field using the same convention for line-type as  in Fig.~\ref{fig:Bfield}a); the emission is particularly efficient and bounds are more effective with respect to the equipartition and reconnection magnetic field cases. At smaller masses the magnetic field matching the synchrotron peak becomes greater than the constant plateau and constraints are quickly relaxed.
The same effect happens for the reconnection magnetic field, at even smaller masses. An analogous effect takes place for the $b -\bar b$ channel, but to a smaller extent due to the soft spectrum.

We have already discussed patterns of dependences of the synchrotron X-ray signal with the magnetic field in many details. For moderate to large values of magnetic fields around the central BH, the limit from the detection of Sgr~A$^*$ by CHANDRA tends to be the tightest in the WIMP parameter space, except if the WIMP mass is too small, the annihilation channel is too soft, or the density of WIMP very close to the  GC is not large enough, i.e. if, in connection to one or more of these issues, we do not have enough high energy radiating electrons and positrons. The signal is generated in a very small region, where the DM profile depends on the ratio $\sigma v/M_\chi$, and hence the scaling of the flux with the cross section is not linear. In the case of the $A_{sp}$ profile, this dependence is so strong that the limit can be double valued.

Finally, the dash-dotted line refers to the limit extracted from detection by CHANDRA of a diffuse X-ray background, when compared to the predicted IC emission on the CMB. It can be the tightest X-ray limit, however, it is never the strongest constraints in any combination of our reference setups.

In general, the request for the WIMP thermal relic abundance to not exceed the value of the mean DM density in the Universe as derived from cosmological measurements, fixes a lower bound on the total annihilation rate at zero temperature (the relic density scales approximately with the inverse of the pair annihilation rate; there are, however, cases when such correspondence is badly violated, the prime example being when coannihilation effects are present). The very tight constraints we have found in case of the $A_{sp}$ profile should make very narrow, or even close, the allowed window in the WIMP parameter space. For the $N_{sp}$ profile, on the other hand, the limits we have derived are much less stringent.

\subsection{Projected constraints with upcoming observations}

Indirect dark matter detection is one of the most ambitious objectives for new observational campaigns or new telescopes getting available in the near future, with the GC often being indicated as the prime observational target. We try to make here a projection on how significant could be the improvement with respect to the region of the WIMP parameter space already excluded in Figs.~\ref{figsvvsmb} and \ref{figsvvsmtau}.

\begin{figure}[t]
 \begin{minipage}[htb]{8cm}
   \centering
   \includegraphics[width=7.8cm]{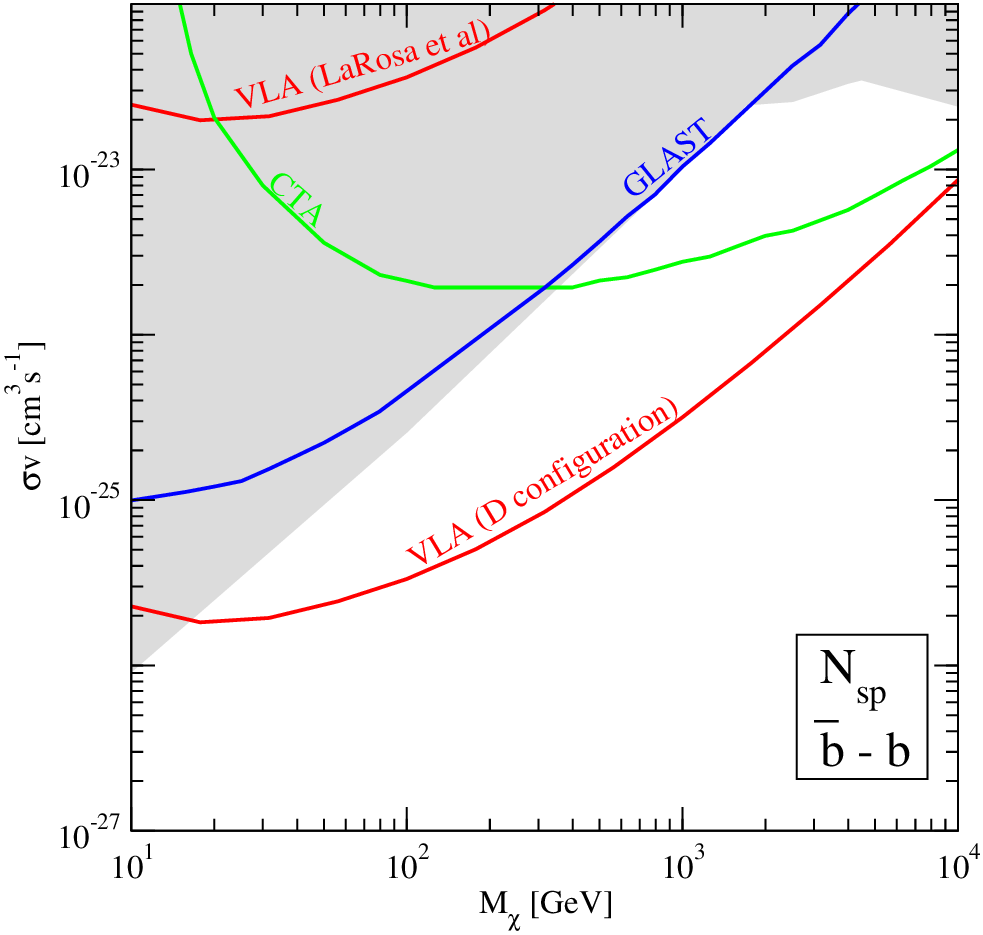}
\end{minipage}
 \ \hspace{2mm} \hspace{3mm} \
 \begin{minipage}[htb]{8cm}
   \centering
   \includegraphics[width=7.8cm]{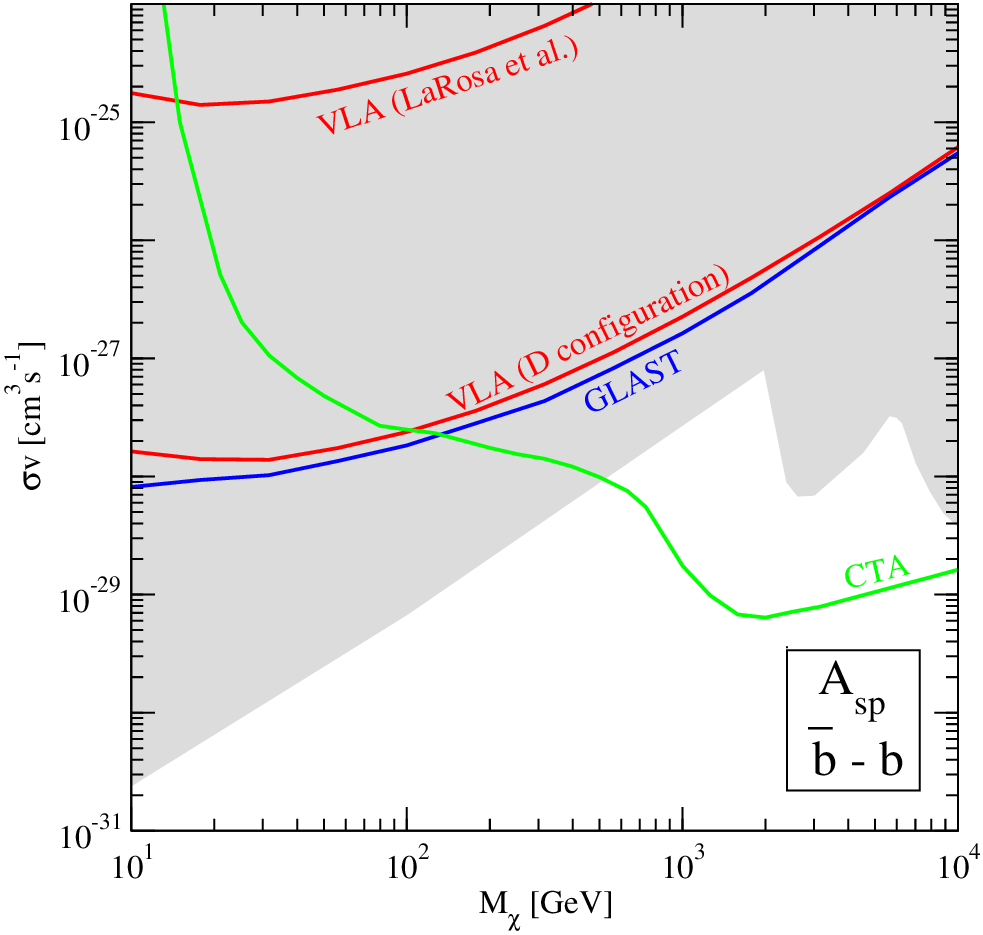}
 \end{minipage}
    \caption{Projected exclusion limits from VLA, GLAST and CTA, in the plane WIMP annihilation cross section versus WIMP mass, in the case of $b-\bar b$ as the dominant annihilation channel.  The {\it Left Panel} and {\it Right Panel} show the limits for, respectively, the $N_{sp}$ and $A_{sp}$ profile. The GLAST and CTA projections are obtained combining an angular and spectral analysis as described in the text. The VLA limit arises from the comparison with the background noise level at 50 arcmin away from the GC. The upper curve is derived assuming a noise level as in \cite{LaRosa:2000}, while the lower curve is computed considering the minimal noise achievable by VLA (D configuration). Shaded regions identify the models violating at least one of the constraints in Fig.~\ref{figsvvsmb} (considering the weakest limit among the three cases with a different choice of the magnetic field radial profile).
}
 \label{figsvvsmprojb}
\end{figure}

\begin{figure}[t]
 \begin{minipage}[htb]{8cm}
   \centering
   \includegraphics[width=7.8cm]{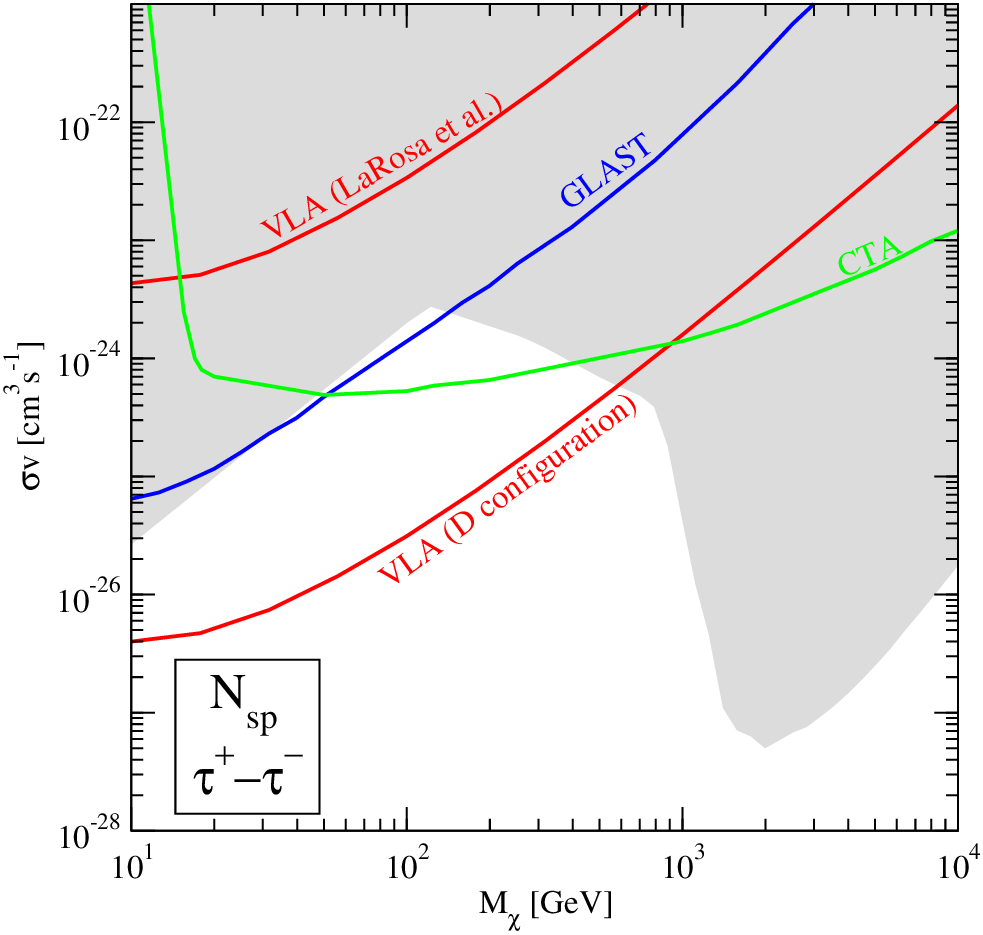}
\end{minipage}
 \ \hspace{2mm} \hspace{3mm} \
 \begin{minipage}[htb]{8cm}
   \centering
   \includegraphics[width=7.8cm]{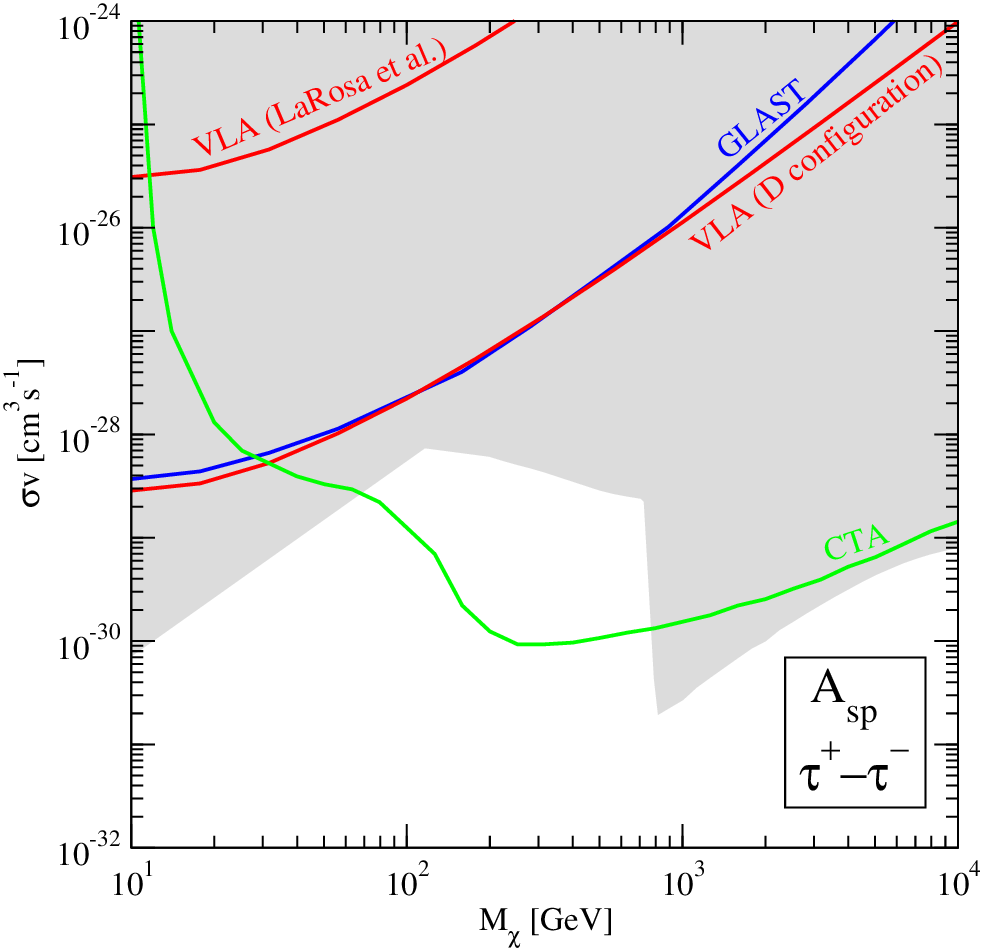}
 \end{minipage}
  \caption{The same as Fig. \ref{figsvvsmprojb}, but taking $\tau^+-\tau^-$ as the dominant annihilation channel.}
 \label{figsvvsmprojtau}
\end{figure}

We mentioned that the radio bounds could become even stronger for wide~field $90$~cm observations of the GC region reaching a noise level which is significantly reduced  with respect to the map constructed in \cite{LaRosa:2000}, at least in case the intrinsic dimension of Sgr~A in the radio band is not much larger than what is inferred from present observations. In Figs.~\ref{figsvvsmprojb} and \ref{figsvvsmprojtau} we sketch the case of a hypothetical observation with the VLA in its configuration with the worst angular resolution for spatial reconstruction, but with the maximal DM-signal to noise ratio, namely with FWHM=$200"$ and a noise level of $0.1$~mJy in $50$~hours of observations (configuration D in Ref.~\cite{VLA})\footnote{The EVLA project~\cite{EVLA}, scheduled for 2013, should improve the continuum sensitivity, and consequently the WIMP constraint, by a factor of 2 to 40 with respect to VLA.}. We are pointing the telescope at an angle of $50'$ with respect to the GC. The lower curves sketch the improvements in upper bounds which could be obtained in the case of presence of regions with no contaminations from astrophysical backgrounds (3~$\sigma$ noise level). This scenario corresponds to the most favorable case. 
Indeed a 90~cm diffuse emission at the GC was already detected~\cite{Haslam:1982,LaRosa:2005ai}. However, the poor angular resolution of the surveys (51 and 39 arcmin, respectively) does not allow to derive the spatial structure of the emission in the innermost region. In the GC image of \cite{LaRosa:2000}, such emission does not seem completely isotropic and hence, from patches of the map with no background, we can extract tighter bounds (upper curves in Figs.~\ref{figsvvsmprojb} and \ref{figsvvsmprojtau}) with respect to \cite{Haslam:1982,LaRosa:2005ai}. Indeed, although the observations made with the Green Bank Telescope and reported in \cite{LaRosa:2005ai}\footnote{Note that the magnetic fields considered in this paper and plotted in Fig.~\ref{fig:Bfield} are consistent with the bound derived in~\cite{LaRosa:2005ai} by the comparison of the detected diffuse non-thermal source and the expected synchrotron emission from GC cosmic rays.} have a comparable sensitivity, the associated image shows a smoother diffuse emission, due to the larger angular resolution, and the comparison between the WIMP signal and the noise level has to be performed at larger angles, where the DM emission is fainter. 
The real limit is probably standing in between the two extreme cases plotted in Figs.~\ref{figsvvsmprojb} and \ref{figsvvsmprojtau}. 

The space satellite GLAST has been launched very recently and will be soon operative. The energy range of detection is
approximately 100~MeV--300~GeV, with an expected sensitivity improved by a factor $100$ with respect to EGRET.
The PSF and the effective area at high energy are respectively $10^{-5}$ sr and $10^4$ cm$^2$ (in the following we will consider the full energy dependence in these quantities as inferred from~\cite{GLAST:2007}; averaging over the angle of observation at which the GC stands with respect to the zenith of the detector are included as well, finding an effective exposure which is essentially reduced by a factor of 0.3). We  have also assumed a $10\%$ energy resolution, an exposure time of $5$ years, and
systematic errors of 5.2$\%$~\cite{GLASTsysterr}; the latter are relevant only at energies $<10$~GeV. 

The next generation of ACT, the Cherenkov Telescope Array (CTA) project, is currently under development. The proposed energy range of detection is 10~GeV--100~TeV, thus overlapping and extending on the HESS range. The most dramatic improvement will be in the effective area, up to about $1$ km$^2$ or even larger in extreme configurations, with highly reduced statistical errors.
Based on the study in~\cite{CTA:2007}, we assume systematic errors to be $\sim$1\%, the energy resolution at the level of $10\%$, and the point spread function equal to $10^{-5}$~sr. For an ACT, on top of astrophysical backgrounds, one needs to take into account the background from misidentified showers, i.e.: 
\be
\frac{dN_{sh}}{dE}= \frac{dN_{had}}{dE}+\frac{dN_{el}}{dE}\label{eqmisshow}
\ee
where $\frac{dN_{had,el}}{dE}$ are the spectra of the gamma-like showers from hadrons and electrons, respectively. We treat these components following~\cite{Bergstrom:1997fj}, assuming $1\%$ of misidentified hadron showers with respect to the total incident flux. We assume a total of about $250$ hours for the exposure time (reasonable in 5 years of operation for CTA).

To estimate the $\gamma$--ray projected constraints in the plane DM mass versus annihilation cross section, we make an extrapolation  of the point--like and diffuse astrophysical backgrounds detected by HESS over the whole energy range of interest, namely 1~GeV--300~GeV for GLAST and 10~GeV--100~TeV for CTA, assuming single power law scaling for the fluxes. 
We consider two generic power law spectra $A_i E_{\gamma}^{-B_i}$, with $i=p,d$, one for the point--like GC source and the other for the diffuse gamma--ray emission in the Galactic center region, assumed to have a flat angular profile. We first generate a sample of $A_i$ and $B_i$ coefficients satisfying the condition $\chi_{red}^2\leq1$ when compared to the data-sets from HESS observations \cite{Aharonian:2004wa} and \cite{Aharonian:2006au}. Then, we simulate how this flux should be seen by GLAST and CTA summing statistical and systematic errors in quadrature (we define the statistical error as the square root of the number of events in any given bin). Finally, we compute the best fits assuming as theoretical flux a dark matter contribution on top of a new two-component background $\tilde A_i E_{\gamma}^{-\tilde B_i}$. Among all the  $\tilde A_i$ and $\tilde B_i$ coefficients allowed, we retain the case providing the smallest $\chi^2$ and take as exclusion criterion $\chi_{red}^2 >3$, namely a flux not well fitted by the dark component plus any viable astrophysical components. The $\chi^2$ analysis is performed both on the energy spectra and on the angular structure of the flux. The angular bin size is fixed according to the PSF.
For the $A_{sp}$ profile, this last step is useless, since the dark matter signal is concentrated in the central angular bin (see Fig. \ref{fig:Iang}), while for the less cuspy $N_{sp}$ profile this procedure provides additional information. (The method we are implementing leads to analogous conclusions with respect to the treatment in~\cite{Dodelson:2007gd}, the main differences in the extrapolated limits stemming from the different halo profiles adopted and a different treatment of systematic errors.)

Results are shown in Figs. \ref{figsvvsmprojb} and \ref{figsvvsmprojtau}. In the same plots, shaded regions identify the models violating at least one of the constraints in Figs.~\ref{figsvvsmb} and \ref{figsvvsmtau} considering the weakest limit among the three cases with a different choice of the magnetic field radial profile, i.e. models that are {\sl excluded} (at least within the rather general set of assumptions we are making regarding magnetic fields, treatment of electrons and positrons propagation, dark matter densities in the GC region, and spectral features of the yield from WIMP annihilations).  The projected  limit for GLAST is always lying in a shaded region; those for CTA span modest portions of the parameter space which are not already excluded. One should consider, on one hand, that we may have been over conservative, since we derived these limits relying on extrapolations on both the energy spectra and the angular profile for the background astrophysical components, as well as without assuming any theoretical modeling of such astrophysical sources; with data at hand the picture may look slightly more favorable. On the other hand, this is indeed suggesting  that, although the $\gamma$-ray band is the regime in which it is most straightforward to make the connection between a given dark matter model and the induced signal, it does not seem to be the energy range with the best signal to background ratios, at least in the case of the GC and of not very cuspy DM profile.

\section{Conclusion}
\label{sec:conclusion}

We have presented a systematic, self-consistent study of the multi--wavelength emission due to WIMP pair annihilations in the Galactic center region. The WIMP signal is expected to extend from the radio 
band up to gamma-ray frequencies. The gamma-ray luminosity is mostly associated to the chain of decays and/or hadronization processes initiated by two-body annihilation channels, leading to the production of neutral pions and their subsequent decays into two photons. In analogous chains, and with comparable efficiencies, high-energy electrons and positrons are produced as well: emitted in a region with large magnetic fields, they give rise to synchrotron emission covering radio frequencies up to, possibly, the X-ray band. A minor role is also played by inverse Compton scattering on the cosmic microwave background or starlight. 

Referring to a generic WIMP DM scenario, we have discussed spectral and angular features,  and  sketched the correlations among signals in the different energy bands. We have illustrated which are the critical assumptions in deriving such conclusions, starting from uncertainties in the DM source functions, regarding both WIMP models and DM distributions, up to the modeling of propagation for electrons and positrons, and the assumptions on magnetic field profiles. We have introduced benchmark cases to guide the discussion and extracted the most relevant general trends: Radio to mm synchrotron emission is essentially independent from the shape of the magnetic field in the innermost region of the Galaxy, while at shorter wavelengths, i.e. in the infrared and, especially, the X--ray band, a different choice for the magnetic field may change predictions dramatically. Radio signals have in general very large angular sizes, larger than the typical size for the source function and hence of the $\gamma$-ray signals. The size of the region of synchrotron X-ray emissivity shrinks dramatically going to larger frequencies, smaller WIMP masses, or softer annihilation channels.

The luminosity of the WIMP source at the different frequencies, and especially comparing the radio to the $\gamma$-ray band, is essentially at a comparable level, with luminosity ratios depending rather weakly on WIMP mass and annihilation channels. This is interesting, since the GC astrophysical source Sgr~A$^*$, an unusual source, certainly very different from typical galactic or extragalactic compact sources associated to black holes, has a very low luminosity over the whole spectrum, at a level at which it is plausible that a WIMP-induced component may be relevant. Indeed, after a closer look, one sees that none of the fluxes detected in GC direction has spectral or angular features typical of a DM source, still all data-sets contribute to place significant constraints on the WIMP parameter space. We have found that, although the $\gamma$-ray band is the regime in which it is most straightforward to make the connection between a given dark matter model and the induced signal (hence it is also the regime on which most of previous analyses have concentrated on),  it does not seem to be the energy range with the best signal to background ratios. In the case of large magnetic fields close to the GC, X-ray data can give much tighter constraints. Radio and NIR measurements, which are less model dependent, tend to be more constraining as well.

Regarding an outlook for the future, we have explored the capability of improving $\gamma$-ray constraints on WIMP models of the GLAST satellite telescope, and of CTA as representative of the next generation of air Cherenkov telescopes. The recent discovery of a $\gamma$-ray GC source and of a diffuse $\gamma$--ray component, however, limits the possibility of dramatic improvements,
possibly reducing the region in the parameter space accessible to $\gamma$-ray telescopes to regimes that, within the range of assumptions listed in our analysis, are already excluded at other wavelengths. On the other hand, if the  Sgr~A source has a size in the radio band which is not significantly larger than its presently estimated value, future wide field radio observations could  be a new effective way to test WIMP DM models.

\section*{Acknowledgements}

We would like to thank G.~Bertone and D.~ Merritt  for kindly providing some of dark matter halo profiles which have been used in the analysis. We also would like to thank S.~Colafrancesco, M. Massardi and A. Michelangeli for useful discussions. 
We thank the anonymous referee for providing comments which helped in improving the present version of the work.
This work is partially supported by the European Community's Human Potential Programme under Contract No. MRTN-CT-2006-035863.

\end{document}